\documentclass[%
 reprint,
 amsmath,amssymb,
 aps,
showkeys
]{revtex4-2}
\usepackage[caption=false]{subfig}
\usepackage{graphicx}
\usepackage{dcolumn}
\usepackage{bm}
\usepackage{xcolor}
\usepackage{blindtext}
\usepackage{hyperref}
\usepackage[markup=underlined]{changes}
\definechangesauthor[name=Ivan, color=blue]{IS}
\definechangesauthor[name=Zikanov, color=red]{OZ}
\usepackage[T2A]{fontenc}			
\usepackage[utf8]{inputenc}
\usepackage[english]{babel}
\usepackage{orcidlink}

\definecolor{reviewer2_color}{RGB}{50,50,150}
\usepackage[markup=underlined]{changes}
\definechangesauthor[color=orange]{Reviewer1}
\definechangesauthor[color=reviewer2_color]{Reviewer2}

\begin{document}

\preprint{APS/123-QED}

\title{
 {Exploratory study of liquid metal response to rapid variation of applied magnetic field} }

\author{Ivan Smolyanov \orcidlink{0000-0001-8224-6976}}
\email{i.a.smolyanov@gmail.com}
\affiliation{Ural Federal University, Ekaterinburg, Russia.}

\author{Oleg Zikanov \orcidlink{0000-0003-3844-1779}}%
 \email{zikanov@umich.edu}
\affiliation{University of Michigan - Dearborn, Dearborn, MI 48128-1491, USA }%

\date{\today}
\begin{abstract}
Transient plasma events, such as plasma disruptions, are anticipated in the future magnetic-confinement nuclear fusion reactors. The events are accompanied by a rapid change in the magnetic field generated by the plasma current and, accordingly, induction of strong eddy currents and Lorentz forces within the reactor structure. This work targets processes within liquid-metal components of the reactor's breeding blankets. Order-of-magnitude analysis and  {exploratory} numerical simulations are performed to understand the response of liquid metal to a rapidly changing magnetic field and to evaluate the accuracy of commonly used simplifying model assumptions. The response is found to consist of two stages: an initial brief stage ($\sim 1$ ms) characterized by a rapid increase in the induced currents, forces, and fluid velocity; and a subsequent stage, which is triggered by the growing velocity of the metal and marked by reversals of Lorentz force, and oscillations and decreases in the amplitude of the induced fields. The transition to the second stage sets the upper limit of the velocity ($\sim 0.5$ m/s in our tests), to which an initially quiescent metal can be accelerated during the event. The simulations indicate that many widely used model assumptions, such as the negligible role of Joule dissipation in the heat balance and the constancy of physical property coefficients, remain valid during the response. However, the assumption of liquid metal incompressibility is found to be questionable due to the potential significant effects of pressure waves.
\end{abstract}

\maketitle


\section{Introduction}\label{sec:intro}
Transient plasma events, such as disruptions, edge-localized modes, or vertical displacements, are anticipated in tokamak fusion reactors operating in a high-performance regime \cite{hollmann2015status}. During any of such events, the state of the plasma changes in a very short time, in the range between $\sim 10^{-4}$ and $\sim 10^{-2}$ s. An important example is the disruption - a thermal quench followed by a quench of the toroidal electric current flowing in the plasma, which occurs on the typical time scale $\sim 10^{-2}$ s. 

In this paper, we are interested in the processes that occur within the blanket - the reactor structure that surrounds the plasma chamber and serves the triple purpose of energy conversion, shielding, and breeding of tritium. More specifically, we consider the processes that occur during and immediately after a disruption event within liquid-metal modules of the blanket. It has been argued that liquid-metal modules, most likely composed of pure Li or Pb-Li alloy, significantly improve the blanket's performance \cite{abdou2015blanket}. 

Since the blanket is separated from the plasma chamber by the first wall, the only direct impact of a plasma transient is via electromagnetic interactions. The interactions are of a unique kind, with nothing similar found in other systems. They are caused by a very rapid change in an exceptionally strong magnetic field imposed in the blanket area. In the case of a plasma disruption, the poloidal magnetic field of strength $\sim 0.5$ T, which is generated in normal operation by the plasma current, completely disappears. This occurs in the presence of a steady toroidal field of strength $\sim 10$ T imposed by the reactor's magnets.

It is evident that such a rapid change of the imposed magnetic field induces strong eddy currents and, correspondingly, strong Lorentz forces in conducting elements of the blanket and other reactor structures. Order-of-magnitude estimates and preliminary simulations indicate electric currents of density $\sim 10^7$ A/m$^2$ and the amplitudes of the Lorentz force exceeding $10^7$ N/m$^3$  \cite{blanchard2019thermal,lei2022study}. The resulting acceleration of the liquid metal (here we estimate for heavier PbLi) can be as high as  $10^3$ m/s$^2$. If applied during the entire duration of the event, this may lead to growth of the metal velocity to tens of m/s. 

The processes within the blanket during a transient plasma event are expected to have a strong impact on the reactor's operation. This may include stresses in the solid structure, strong mixing in the liquid-metal components, and the electromagnetic effect on the plasma behavior. Hypothetically, these effects could significantly modify the course of the event itself.

 {The response of liquid metal to a rapidly changing magnetic field at the parameters relevant to the blanket environment was a subject of several earlier works. Attempts of numerical simulations of an entire blanket or its module have been made (see, e.g., \cite{lei2022study,rozov2014strategy}), but they inevitably use oversimplified models and suffer the effects of insufficient numerical resolution. Some insight into the physics of the response was provided in the studies of the geometries of an infinite channel \cite{kawczynski:2018thesis} and infinite duct \cite{smolentsev:2025}. In both cases, the liquid metal was assumed to be an incompressible fluid. The imposed magnetic field had two components, a constant transverse component and a longitudinal component decaying at a constant rate. Steady-state one-dimensional (in \cite{kawczynski:2018thesis}) and steady-state and transient two-dimensional (in \cite{smolentsev:2025}) solutions were found. As will become clear in the following, these model settings are analogous to ours, except that in our case the duct has a finite length, solutions are three-dimensional and inherently unsteady, and a more realistic time variation of the longitudinal field is assumed. We will see that there are certain qualitative similarities between our results and the predictions of \cite{kawczynski:2018thesis,smolentsev:2025}. }

The nature and extent of the blanket response to transient plasma events are far from clear at the moment. One major reason for this is the physical complexity of the process, which makes it challenging to model accurately. Furthermore, no experimental studies that approximate the conditions of a fusion reactor are currently feasible.  

Yet another challenge to understanding the blanket response is the lack of clarity on applicability of the various commonly applied simplifying assumptions. Can the unique physical effects that appear during a transient plasma event invalidate some of the assumptions accepted to be accurate in the analysis of steady-state operation?

This paper presents the results of an exploratory study of the processes caused in a liquid-metal module of a breeding blanket during a transient plasma event. We begin by examining the order-of-magnitude analysis of the typical scales and a general discussion of validity of the commonly used assumptions of liquid-metal behavior (see Section \ref{sec:models}) Further analysis based on the quantitative data generated by numerical simulations of a simplified system is presented in sections \ref{sec:method} and \ref{sec:results}. 

\section{Applicability of commonly used models}\label{sec:models}
\subsection{Models commonly applied to liquid metal flows in steady-state operation of blankets}
In the steady-state operation of a recator, liquid-metal components of a breeding blanket exist under unique conditions created by an exceptionally strong imposed magnetic field and high energy flux. This severely limits the usefulness of the predictions based on extrapolation of the flow behaviors known for more conventional settings. Experimental studies representing the real reactor environment are impossible until a full-scale  reactor with a fully functional blanket begins to operate. This makes the numerical simulation a primary approach to the analysis. This is a relatively mature area (see, e.g., \cite{mistrangelo:2021a,mistrangelo:2021b} for recent reviews and \cite{chen2020toward,chen2022toward} for recent large-scale simulations). 

The numerical models applied to liquid metal flows in blankets are based on a virtually universally adopted set of simplifying assumptions justified by theoretical arguments and the typical values of parameters \cite{mistrangelo:2021a,mistrangelo:2021b,chen2020toward,chen2022toward,smolentsev2015approach}. The metal is modeled as a single-phase liquid with constant physical properties. The density is assumed constant, unless the thermal convection effect included in a non-isothermal analysis, in which case the density in the buoyancy force term is assumed to be a linear function of temperature according to the Oberbeck-Boussinesq approximation. Heat generation by the Joule and viscous dissipations is neglected in the energy balance. The quasi-static (inductionless) approximation is applied to the magnetohydrodynamic (MHD) effects caused by the imposed steady magnetic field (see, e.g., \cite{davidson2016} for a derivation and discussion).  The approximation uses the fact that the perturbation magnetic field associated with the electric currents induced in a moving liquid metal is much weaker than the imposed field and, therefore, can be neglected in the expressions for Ohm's law and Lorentz force. Furthermore, the typical time of the diffusion of the magnetic field is much smaller than the time scales of the flow, so the induced magnetic field can be approximated as adjusting instantaneously to changes of velocity field. 

\subsection{Applicability to the case of a rapidly changing magnetic field}
\emph{Which, if any, of the model assumptions commonly adopted in the  analysis of liquid-metal flows in a steady-state blanket operation become invalid in the situation where the imposed magnetic field is rapidly changed by a transient plasma event?} The ultimate answer to this question will require extensive and, likely, challenging simulations and experiments.  Preliminary conclusions can, however, be achieved in an order-of-magnitude analysis of the typical amplitudes of the electric current, Lorentz force, and Joule dissipation, and the typical length and time scales of the process. The analysis is presented in this section. 

To be specific, we will use the properties of the eutectic PbLi alloy (currently considered the most likely candidate for liquid metal blanket modules) at 573 K shown in Table \ref{table1}. 

The typical length scale $L$ of the problem is the size of a blanket module filled with liquid metal. The geometry of a module varies depending on the design of the blanket but in a simplified way sufficient for the purposes of our study can be considered as that of a cuboid extended in the poloidal or toroidal direction (in the reactor coordinates) \cite{abdou2015blanket,mistrangelo:2021a,mistrangelo:2021b}. The shorter side of the cuboid is 10-20 cm, while the longer side can be up to 2 m. We will use two typical length scales: $L\sim 0.1$  m corresponding to the radial dimension (roughly along the distance from the plasma chamber) of the cuboid, and $L\sim 1$ m for the longer (poloidal or toroidal) side. 

The typical time scale of the problem $\tau$ is the typical time of the variation of the magnetic field imposed in the blanket area. The variation is caused by a transient plasma event, whose duration varies from $\sim 10^{-4}$ s for edge localized modes to $\sim 10^{-2}$ s for disruptions or vertical displacements. Evaluating $\tau$, we have to take into account the diffusion of the magnetic field within the blanket structure in the radial direction, that is, on the length scale $L_m\sim 0.1$ m.  The typical diffusion time scale is $\tau_m\sim L_m^2/\eta$, where $\eta$ is the magnetic diffusivity coefficient listed in table \ref{table1}. $\tau_m$ can be taken as the lower limit for the magnetic field variations within the blanket, so $\tau\sim 10^{-2}$ s is assumed below.

\begin{table}
    \centering
    \begin{tabular}{l | c | c }
        Property & Value & Ref. \\
        \hline
      Density $\rho$ & $9.4\times 10^3$ kg/m$^3$ & \cite{zikanov2021mixed} \\
      Electric conductivity $\sigma$ & $0.79\times 10^6$ 1/$\Omega$ m & \cite{zikanov2021mixed} \\
      Kinematic viscosity $\nu$ & $2.1\times 10^{-7}$ m$^2$/s & \cite{zikanov2021mixed} \\
      Magnetic diffusivity $\eta$ & 0.993 m$^2$/s & \cite{zikanov2021mixed}\\ 
      Thermal conductivity $\lambda$ & 13.2 W/m K  & \cite{zikanov2021mixed} \\
     Thermal expansion coeff. $\alpha$ & $0.9\times 10^{-4}$ 1/K  & \cite{zikanov2021mixed} \\
      Specific heat $C_p$ & 183 J/kg K & \cite{zikanov2021mixed} \\
       Speed of sound $C$ & 1784 m/s & \cite{ueki2009acoustic} 
    \end{tabular}
    \caption{Properties of PbLi eutectic alloy at 573 K used in the analysis of typical scales.}
    \label{table1}
\end{table}

We start with the applicability of the quasi-static approximation of MHD effects. The formal conditions are that  the magnetic Prandtl and Reynolds numbers are both small \cite{davidson2016}:
\begin{equation}\label{PmRm}
    P_m\equiv \frac{\nu}{\eta}\ll 1, \:\: Re_m\equiv \frac{UL}{\eta}\ll 1,
\end{equation}
where $U$ is the typical velocity of the flow. For the parameters in table \ref{table1}, $P_m\approx 2.1\times 10^{-7}$ and $Re_m\approx (1.007 s/m^2) (UL)$. $P_m$ is very small, but $Re_m$ can be $\sim 1$ or, at the velocity of the liquid metal reaching tens m/s, significantly greater than 1. 

The condition $P_m\ll 1$ implies that the magnetic diffusion time $\tau_m$ is much shorter than the viscous diffusion time $\tau_{\nu}\equiv L^2/\nu$ of the flow. In the conventional analysis, this justifies the assumption that the magnetic field adjusts instantaneously to flow changes. This logic is inapplicable in our case, since $\tau_m$ must be compared with the time scale of the event $\tau$ rather than with the typical time scales of the flow. As we have evaluated, $\tau_m$ is of the same order of magnitude as $\tau$. 

If $Re_m\ll 1$, the effect of the fluid velocity on the induced magnetic field is much weaker than the effect of the magnetic diffusion. The field-velocity coupling term in the induction equation 
\begin{equation}\label{magindeq}
    \frac{\partial \mathbf{B}}{\partial t} = \nabla\times (\mathbf{u}\times \mathbf{B})+\lambda \nabla^2\mathbf{B}
\end{equation}
can be dropped. Since the condition is generally not satisfied in our case, such an approximation would be illegal. The numerical simulations reported later in this paper confirm the significant impact of the velocity-field coupling on the evolution of the magnetic field.

We will now consider the simplifications typically made in models of the hydrodynamics of liquid-metal flows. The first to be considered is the incompressibility approximation, which is valid if the magnitude of the density variation $\Delta \rho$ is consistently much smaller than the reference (e.g., average) density $\rho_0$:
\begin{equation}\label{incomp}
    \frac{\Delta \rho }{\rho_0}\ll 1.
\end{equation}
For the density variations caused by a flow, (\ref{incomp}) is equivalent to the condition on the square of the Mach number
\begin{equation}\label{mach}
    M^2\equiv \left(\frac{U}{C}\right)^2\ll 1,
\end{equation}
where $C$ is the speed of sound listed in table \ref{table1}. It is satisfied at $U$ up to a few hundred m/s. 

Variations of density may also be caused by the direct action of a spatially non-uniform Lorentz force. In order to assess this effect, we calculate the compressibility coefficient using the data in table \ref{table1}. For the isentropic compressibility, we have
\begin{equation}\label{beta}
    \beta_s\equiv \frac{1}{\rho_0}\left(\frac{\partial \rho}{\partial p}\right)_s=\frac{1}{\rho_0 C^2}\approx 3.34\times 10^{-11} \text{Pa}^{-1}.
\end{equation}
The isothermal compressibility is only slightly different:
\begin{equation}\label{betat}
    \beta_T\equiv \frac{1}{\rho_0}\left(\frac{\partial p}{\partial \rho}\right)_T = \beta_s+\frac{\alpha^2 T}{\rho_0 C_p}\approx 3.61\times 10^{-11} \text{Pa}^{-1}.
\end{equation}
Using \eqref{beta} and evaluating the gradient of pressure via the balance with the Lorentz force of the typical amplitude $F$, so $\Delta p\sim L F$, we obtain the estimate
\begin{equation}\label{dp}
    \frac{\Delta \rho}{\rho_0} \sim \beta_s \Delta p\sim \beta_s L F\approx  \left(3.34 \times 10^{-11} \frac{\text{m}^3}{N}\right) F,
\end{equation}
for  $L=1$ m and a tenth of that at $L=0.1$ m. As an example, with preliminary estimates $F\sim 10^7$ N/m$^3$, this formula gives us negligibly small density variations $\Delta \rho < 10^{-3}\rho_0$. 

Yet another component of the incompressibility model is that the propagation of the variations of pressure is assumed to occur at infinite speed (the asymptotic limit of zero Mach number). In our case, this assumption is formally correct in the sense that (\ref{mach}) is satisfied. However, in reality, the assumption is invalid because the time scale $\tau_f\equiv L/U$ based on the fluid velocity is irrelevant. The time of the propagation of pressure waves across the domain $\tau_w\equiv L/C$ must be compared with the typical time $\tau_L$ of variation of the Lorentz force during the event, rather than with $\tau_f$. Considering the worst-case scenario, in which the pressure wave propagates along the longer side of a blanket module, we take $L=1$ m and find $\tau_w\approx 5\times 10^{-4}$ s. Pressure waves may need to be considered and, thus, the hydrodynamic equations of a compressible fluid solved if $\tau_w$ is not much smaller than $\tau_L$.

The volumetric density of the rate of heat generation by the Joule dissipation is estimated as 
\begin{equation}\label{joule}
    \mu\sim \sigma^{-1} J^2\approx (1.27\times 10^{-6} \Omega m)J^2,
\end{equation}
where $J$ is the typical amplitude of the density of induced electric currents. As an illustration, we take the preliminary estimate $J\sim 10^7$ A/m$^2$ to find $\mu\approx 1.3$ MW/m$^3$. This is of the same order of magnitude as the steady-state rate of energy deposition within a blanket \cite{abdou2015blanket}. Since the energy input (\ref{joule}) is active only during the short time $\tau$, its impact on the total energy balance is negligible. 

We can also use the estimate of $\mu$ to assess the validity of the assumption of constant physical properties of the metal. The only plausibly possible cause of a significant variation of the properties during a transient plasma event is their dependence on temperature. As an example, the local increase of temperature of the liquid metal generated by $\mu\approx 1.3$ MW/m$^3$ over  $\tau\approx 10^{-2}$ s is:
\begin{equation}\label{temp}
    \Delta T\sim \frac{\mu\tau}{\rho_0 C_p}\approx 0.7 \text{ K}.
\end{equation}
The variation of the coefficients of the physical properties of the metal, such as those in table \ref{table1}, over such a temperature range is negligibly small. The situation may change if the electric currents are stronger. The specific assessment varies depending on the metal and the coefficient under consideration. For the PbLi alloy, the strongest variation is observed for the dynamic viscosity $\nu \rho$ and the thermal conductivity $\lambda$, which vary by, respectively, $\sim 10$\% and $\sim 5$\% at $\Delta T\sim 50$ K (see \cite{zikanov2021mixed} or \cite{de2008lead} for a review and references to databases). We conclude that the coefficients can be considered constant unless currents of density $J\sim 10^8$ A/m$^2$ or greater are generated in the course of the event. 

The estimates presented in this section can serve as a starting point for answering the question posed at the beginning. We see that the quasi-static (inductionless) approximation is definitely invalid in the situation of a rapidly changing imposed magnetic field. The complete MHD problem, for example, in the form of the equation \eqref{magindeq} must be solved. The validity of the assumptions of incompressibility, constant physical properties, and negligible effect of Joule dissipation on the energy balance can be valid or not depending on the magnitude of the induced electric current $J$ and the Lorentz force $F$. 

Numerical simulations designed to provide the quantitative data necessary for the analysis are presented in sections \ref{sec:method} and \ref{sec:results} of this article.

\section{Approach to numerical analysis}\label{sec:method}
\subsection{Setting}\label{sec:setting}
\begin{figure}[h]
    \centering
    \includegraphics[width=1.0\linewidth]{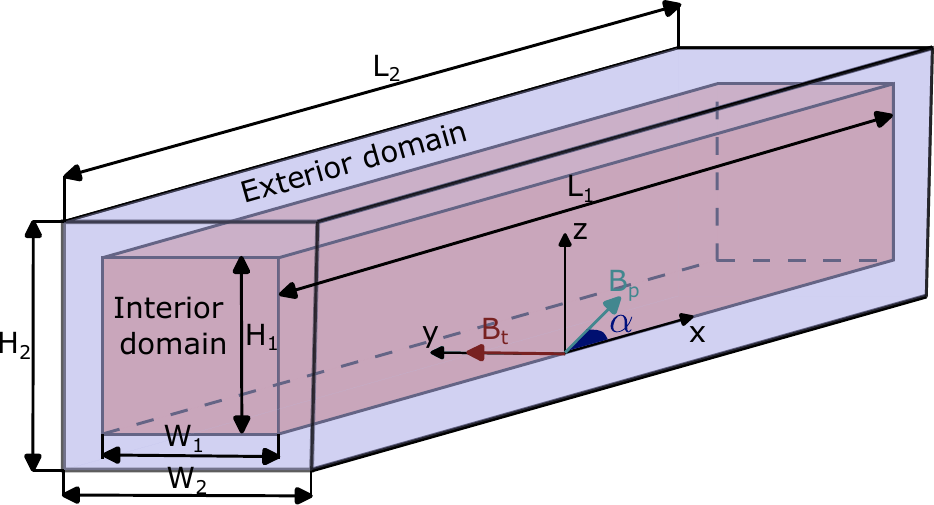}
    \caption{Model geometry used in numerical simulations. The interior domain of dimensions $L_1\times W_1\times H_1$ is filled with an electrically conducting material (a liquid metal) and has electrically perfectly insulating boundaries. The exterior domain of dimensions $L_2\times W_2\times H_2$ includes a non-conducting extension used for calculations of the induced magnetic field. Components $\mathbf{B}_t$ and $\mathbf{B}_p$ of the imposed magnetic field are shown. }
    \label{fig:Sketch}
\end{figure}

Numerical simulations are performed for the model system shown in figure \ref{fig:Sketch}. The interior domain is a cuboid cavity filled with PbLi at 573 K, whose physical properties are listed in table \ref{table1}.  The cavity has a square cross section with the side $W_1=H_1=0.1$ m and length $L_1=1$ m. The exterior domain is a larger cuboid cavity, which includes the interior domain and an electrically non-conducting extension needed for calculations of the induced magnetic field.

The system is a simplified model of a liquid-metal module of a fusion reactor blanket. For specificity, we focus on the popular type of blanket design, in which the liquid metal circulates through multiple poloidally oriented ducts (e.g., the DCLL blanket design \cite{abdou2015blanket}). The interior domain represents one of such ducts. The coordinates $x$, $y$, and $z$ (see figure \ref{fig:Sketch}) correspond, respectively, to the local poloidal, toroidal, and radial directions of the standard reactor coordinate system. 

The liquid metal is initially at rest, but is set to motion at $t>0$ by the variation of the magnetic field. This is an idealization of the actual situation, in which a liquid metal flow of typical velocity between $\sim 1$ mm/s and a few tens of cm/s, depending on the type of the blanket, is generated during steady-state operation by an imposed pressure gradient or thermal convection effects \cite{abdou2015blanket}.

The imposed magnetic field $\mathbf{B}_0$ has a constant toroidal component $\mathbf{B}_t=B_t\mathbf{e}_y$ and a poloidal component $\mathbf{B}_p$, which in our scenario is a function of time $t$. We will also consider the factor of imperfect alignment between the duct and the poloidal magnetic field in the poloidal-radial plane. which is expected in actual reactor blankets. This can be described by a small non-zero angle $\alpha$ between $\mathbf{B}_p$ and the $x$-axis. The imposed magnetic field is, thus, defined as
\begin{equation}
    \mathbf{B}_0 \left(\mathbf{x}, t \right) = (B_p(t) \cos{\alpha} )\mathbf{e}_x+B_t \mathbf{e}_y+(B_p(t)\sin{\alpha} )\mathbf{e}_z.
    \label{eq:B_ext}
\end{equation}
The model used for the magnitudes $B_t$ and $B_p$ represents, in a simplified manner, a plasma-disruption event in a future full-scale tokamak reactor \cite{abdou2015blanket,blanchard2019thermal}. The typical expected values in steady-state operation are $B_t=10$ T, $B_p^0=0.5$ T. The effect of the plasma disruption is modeled as a decrease of $B_p$ starting at $t=0$ and following
\begin{equation}\label{eq:bp_exp}
    B_p(t) = B_p^0  \exp{(-t/\tau_0)},
\end{equation}
where $\tau_0=6\times 10^{-3}$ s. 

We must note that \eqref{eq:bp_exp} is an idealization. The actual behavior of $B_p$ is more complex and specific to the design of the reactor and the electromagnetic model applied to calculate it \cite{hollmann2015status,blanchard2019thermal,lei2022study,rozov2014strategy}. The approximation \eqref{eq:bp_exp} is sufficient for the purposes of our exploratory analysis. The exponential decay is consistent with the diffusive propagation of the magnetic field perturbation into the blanket. The constant $\tau_0$ determines the time scale of the event. In our case, $B_p$ decays to 1\% of its initial value at $t\approx 28$ ms.

\subsection{Model}\label{sec:model}
The objective of the simulations is to obtain the induced magnetic field $\mathbf{b}(\mathbf{x},t)$ and the electric current density $\mathbf{J}(\mathbf{x},t)$ that appear in the interior domain in response to the varying imposed magnetic field \eqref{eq:bp_exp}.  This will allow us to compute  the Lorentz force and Joule dissipation
\begin{eqnarray}
    \mathbf{F}  & = & \mathbf{J} \times \mathbf{B}, \label{eq:LorentzForce}\\
    \mu & = & \sigma^{-1}|\mathbf{J}|^2,
\end{eqnarray}
where $\mathbf{B}=\mathbf{B}_0+\mathbf{b}$.

Full non-relativistic finite-resistivity MHD model is applied for the electromagnetic fields (see, e.g., \cite{davidson2016}):   
\begin{eqnarray}
    \label{eq:ampere} \nabla\times \mathbf{B} & = & \nabla\times \mathbf{b} = \mu_0 \mathbf{J},\\
    \nabla\cdot \mathbf{J} & = & 0,\\
    \label{eq:faradey} \nabla \times \mathbf{E}_i & = & -\frac{\partial \mathbf{B}}{\partial t}, \\
    \nabla\cdot \mathbf{B} & = & 0,\\
   \label{eq:ohm} \mathbf{J} & = & \sigma \left(\mathbf{E}_i + \mathbf{u} \times \mathbf{B} \right),
\end{eqnarray}
where $\mu_0$ is the absolute magnetic permeability in vacuum, $\mathbf{E}_i$ is the induced electric field and $\mathbf{u}$ is the velocity of the conducting medium in the interior domain. The electrostatic component of the electric field does not appear in \eqref{eq:ohm}, since the standard MHD approximation assumes a negligible effect of free charges. 

 {The entire set of the MHD equations \eqref{eq:ampere}-\eqref{eq:ohm} is shown here in order to facilitate the discussion. The actual simulations use the model based on } the representation in terms of the magnetic vector potential $\mathbf{A}$, such that $\mathbf{B} = \nabla \times \mathbf{A}$ and $\mathbf{E}_i=-\partial \mathbf{A}/\partial t$. For convenience, the potential is separated into components corresponding to the induced and imposed magnetic field: $\mathbf{A}=\mathbf{A}_0+\mathbf{a}$, where $\mathbf{B}_0=\nabla\times \mathbf{A}_0$ and $\mathbf{b}=\nabla\times \mathbf{a}$. The potential satisfies

\begin{equation}\label{eq:comsol:A-formulation}
    \nabla \left(\nabla \cdot \mathbf{a} \right) - \nabla^2 \mathbf{a} = \mu_0 \sigma \left[ -\frac{\partial \mathbf{A}}{\partial t} + \mathbf{u} \times \nabla \times\mathbf{A}\right].
\end{equation}
 {The electric currents  $\mathbf{J}$ are calculated using Ohm's law \eqref{eq:ohm}.}

Two versions of the electromagnetic boundary conditions are used. One is the simplest option available. We solve \eqref{eq:comsol:A-formulation} only within the liquid-metal cavity (the interior domain in figure \ref{fig:Sketch}) and impose the so-called pseudo-vacuum boundary condition on the induced magnetic field
\begin{equation}\label{eq:PSVBC}
    \mathbf{n} \times \mathbf{b} = \mathbf{n} \times (\nabla \times \mathbf{a})  = 0 \text{ at boundary},
\end{equation}
where $\mathbf{n}$ is the normal to the boundary.  {The pseudo-vacuum boundary condition reduces to a Dirichlet condition when expressed via magnetic flux density, but becomes a Neumann-type condition when formulated for the magnetic vector potential.} The condition implies that the tangential component $\mathbf{b}_{\tau}$  and the normal component $J_n$ are zero at the boundary. Since $\nabla\cdot \mathbf{b}=0$, \eqref{eq:PSVBC} also implies $\partial b_n/\partial n=0$ at the boundary.

Condition \eqref{eq:PSVBC} is often applied in exploratory or benchmark solutions when the tangential component of the magnetic field at the boundary of a conducting region can be assumed to be zero (see, e.g., \cite{jackson2014spherical}) or in MHD problems at moderate magnetic Reynolds numbers, when $|\mathbf{b}|\ll| \mathbf{B}_0|$  (see, e.g., \cite{bandaru2016hybrid}). However, in the general case, the validity of the pseudo-vacuum condition \eqref{eq:PSVBC} cannot be assumed a priori.  {A more detailed discussion can be found, e.g., in \cite{bandaru2016hybrid,birk2022magnetostatic,kawczynski:2018thesis}.}

A more accurate and broadly applicable solution must include the induced magnetic field both inside and outside the conducting domain. This is done in the second version of the boundary conditions used in our study. The domain in which \eqref{eq:comsol:A-formulation} is solved is extended on all sides by a finite non-conducting buffer zone (the exterior domain in figure \ref{fig:Sketch}). Conditions \eqref{eq:PSVBC} are imposed at the outer boundary of the exterior domain. Due to the rapid decay of $\mathbf{b}$ with distance from the conducting zone, an accurate approximation by this approach requires a moderately large extension.  {Similar domain extension techniques was employed in \cite{birk2022magnetostatic}. A slightly different version with the true far-field condition $\bm{b}=0$ imposed at the outer boundary of the extended domain was used in \cite{kawczynski:2018,kawczynski:2018thesis, bandaru2016hybrid}}. Our test calculations show that an extension to $(L_2,W_2,H_2)\ge 3(L_1,W_1,H_1)$  is sufficient. A further increase in $L_2$, $W_2$, or $H_2$ does not affect the solution within the interior domain. 
    
An approximation is adopted for the hydrodynamic part of the problem. The velocity field $\mathbf{u}(\mathbf{x},t)$ of liquid metal in the interior domain is calculated in the course of the solution using the momentum equation in the form 
\begin{equation}\label{eq:velocity}
    \rho \frac{\partial \mathbf{u}}{\partial t} = \mathbf{F}(\mathbf{x},t), 
\end{equation}
where $\mathbf{F}$ is the Lorentz force \eqref{eq:LorentzForce}. The other hydrodynamic effects:  {mass conservation of the fluid}, viscosity, advective transport,  {pressure force}, and boundary conditions on velocity at solid walls, are ignored. Such a curtailed modeling approach is justified by the focus of our study on the immediate (on the time scale $\tau\sim 10^{-2}$ s) response of the liquid to the strong and unsteady Lorentz force. The neglected hydrodynamic phenomena appear as a consequence of this response, but evolve on larger typical time scales, such as the advection $\tau_a\sim L/U$ and viscous $\tau_{\nu}\sim L^2/\nu$ scales. Another probably more important argument is that at the current level of understanding, including the full hydrodynamic behavior would have to be based on an \emph{arbitrarily chosen} hydrodynamic model. As we have discussed previously, selection of the model is not self-evident and requires that the induced electric currents and the forces acting on the liquid are evaluated first. 

Two modeling approaches are tested to explore the effect of fluid acceleration on the dynamics of electromagnetic fields. In one,  $\mathbf{u}$ is neglected in Ohm's law \eqref{eq:ohm}. In the other,  $\mathbf{u}$ is included in Ohm's law, so it appears in the expression on the right-hand side of \eqref{eq:comsol:A-formulation} and affects the Lorentz force. 

\begin{table}    
    \centering
    \begin{tabular}{l | c | c }
    Model & Boundary conditions &  Velocity in Ohm's law\\ 
    \hline
    1 & Pseudo-Vacuum & No \\
      2 & Extended domain & No \\
     3 & Pseudo-Vacuum & Yes \\
      4 & Extended domain & Yes \\
    \end{tabular}  
    \caption{Models used in simulations. See text for explanation.}
    \label{table_model}
    \end{table}

To evaluate the increase in temperature of the liquid caused by the Joule dissipation of the induced currents, the heat equation 

\begin{equation}\label{eq:ht}
	\rho C_p \frac{\partial T}{\partial t} 
 = \frac{|\mathbf{J}|^2}{\sigma}
\end{equation}
is solved in the interior domain. The terms for heat conduction and convection are omitted in (\ref{eq:ht}) because their typical time scales $\tau_a\sim L/U$ and $\tau_{\chi}\sim L^2\rho C_p/\lambda$ are substantially larger than the time scale of the effects considered here.

In order to understand the influence of the modeling assumptions on the solution,  simulations at $\alpha=0$ are repeated using four approaches (see  table \ref{table_model}). In the simplest model 1, \eqref{eq:comsol:A-formulation} and \eqref{eq:ht} are solved in the interior domain with the boundary conditions \eqref{eq:PSVBC}  imposed at the boundary of that domain. The velocity $\mathbf{u}$ remains zero during the simulations.  Model 2 presents an improved approximation of the magnetic field, such that  \eqref{eq:comsol:A-formulation} is solved in the extended (interior plus exterior) domain, and the magnetic field boundary conditions \eqref{eq:PSVBC} are imposed at the outer boundary of the exterior domain. The acceleration of the liquid is taken into account in model 3. In this model,  \eqref{eq:comsol:A-formulation}, \eqref{eq:velocity}, and \eqref{eq:ht} are solved in the interior domain with the boundary conditions \eqref{eq:PSVBC}.  Finally, model 4 includes both modifications: the extended domain and nonzero fluid velocity. In this model,  \eqref{eq:comsol:A-formulation} is solved in the extended domain (interior plus exterior), while \eqref{eq:velocity} and \eqref{eq:ht} are solved in the interior domain.

In order to understand the energy aspects of the process, we calculate the kinetic and internal energies gained by the fluid
\begin{eqnarray}
    \label{eq:ke} W_k & = & \int_{V_I} \rho \frac{|\mathbf{u}|^2 }{2}dV,\\
    \label{eq:ie} W_i & = & \int_{V_I} \rho C_p(T-T_0) dV,\\
\end{eqnarray}
where $V_I$ is the volume of the interior domain and $T_0=573$ K is the reference temperature. The energy of the magnetic field is separated into the imposed and induced components:
\begin{equation}
    \label{eq:me} W_m=W_m^0+W_m^i, \: W_m^0=\int_{V_E} \frac{|\mathbf{B}_0|^2}{2\mu_0}, \: W_m^i=\int_{V_E} \frac{|\mathbf{b}|^2}{2\mu_0},
\end{equation}
where $V_E$ stands for the extended domain.

The energy balance is
\begin{eqnarray}
    \label{eq:balme} \frac{d W_m^i}{dt} & = &-\frac{dW_m^0}{dt}-Q_h-Q_L-S,\\
    \label{eq:balke} \frac{d W_k}{dt} & = & Q_L,\\
    \label{eq:balie} \frac{d W_i}{dt} & = & Q_h,\\
\end{eqnarray}
where 
\begin{eqnarray}
\label{eq:qh}    Q_h & = & \int_{V_I} \frac{|\mathbf{J}|^2}{\sigma} dV,\\
  \label{eq:ql}   Q_L &= & \int_{V_I} (\mathbf{J}\times \mathbf{B})\cdot \mathbf{u} dV
\end{eqnarray}
are the total rates of the Joule dissipation and mechanical work of the Lorentz force and
\begin{equation}
    \label{eq:pf} S =\int_{\partial V_E} \frac{\mathbf{E}_i\times \mathbf{B}}{\mu_0} \cdot \mathbf{n} dS
\end{equation}
is the Poynting energy flux through the boundary of the extended domain.

\subsection{Numerical method}\label{sec:num_method}
Each simulation starts at $t=0$ with the initial conditions $\mathbf{a}=0$, $\mathbf{u}=0$, $T=573$ K. The solution is computed at $0<t\le t_{final}=45\times 10^{-3}$ s with the imposed magnetic field varying according to \eqref{eq:B_ext}, \eqref{eq:bp_exp}.

The finite element solver COMSOL Multiphysics® is utilized \cite{comsol}. The A-formulation of the Maxwell equations in the magnetic-field interface is adapted to match the models described in section \ref{sec:model}. 

A rectangular and structured computational grid is used. In the interior domain, the grid is clustered near the walls. In each direction the wall-normal cell size increases from the boundary to the center of the domain according to
\begin{equation}\label{eq:mesh_size}
    h_n = h_1\left[1 + \frac{\left(n-1 \right) \left( k_{rate} - 1\right)}{N-1} \right],
\end{equation}
where $n \in [1, 2, ..., N]$ is the local index of a cell, $N$ is the total number of cells between the wall and the center, and $k_{rate}$ is the clustering coefficient defined as the ratio between the maximum $h_N$ and minimum $h_1$ sizes. If models 2 and 4 of table \ref{table_model} are used, the exterior domain is discretized by the grid with 6 cells in each wall-normal direction, with the wall-normal size growing outwards according to \eqref{eq:mesh_size} with the same $k_{rate}$ as in the interior domain.

Spatial discretization is based on linear shape functions. The grid sensitivity study has determined that the interior grid with $N_x\times N_y\times N_z = 100\times 30 \times 30$ cells and $k_{rate}=6$ is sufficient for accurate discretization.

The time discretization is by the implicit backward differentiation formula with variable discretization order and time step. The discretization order varies between 2 and 5 according to the adaptive algorithm of COMSOL Multiphysics® \cite{comsol}. The upper limit on the allowed time steps is set manually by the formula for time layers $t_i$
\begin{equation}\label{eq:time}
  t_i = \left[\frac{\tanh\left(D \left( \frac{i-1}{N_t-1} - 1 \right)\right)}{\tanh(D)} + 1\right]t_{final}, \: i=1,\ldots,N_t,
\end{equation}
where $D=1.5$ and $N_t=300$. According to \eqref{eq:time}, smaller time steps are utilized in the early stage of the solution, which is characterized by a particularly rapid variation of variables. The limits on the smallest and the largest time steps are 0.0453 ms and 0.2494 ms, respectively. Further refinement of the time steps is provided by the time-step adjustment tool of COMSOL Multiphysics® \cite{comsol,soderlind2006adaptive}. The tool ensures that the estimated local truncation error does not exceed the relative tolerance of $10^{-6}$. 

The time layers \eqref{eq:time} are also used to save the solution variables for further analysis.

\section{Results}\label{sec:results}

\subsection{Model comparison}\label{sec:mod_comp}
\begin{figure}
    \centering
    \includegraphics[width=1.0\linewidth]{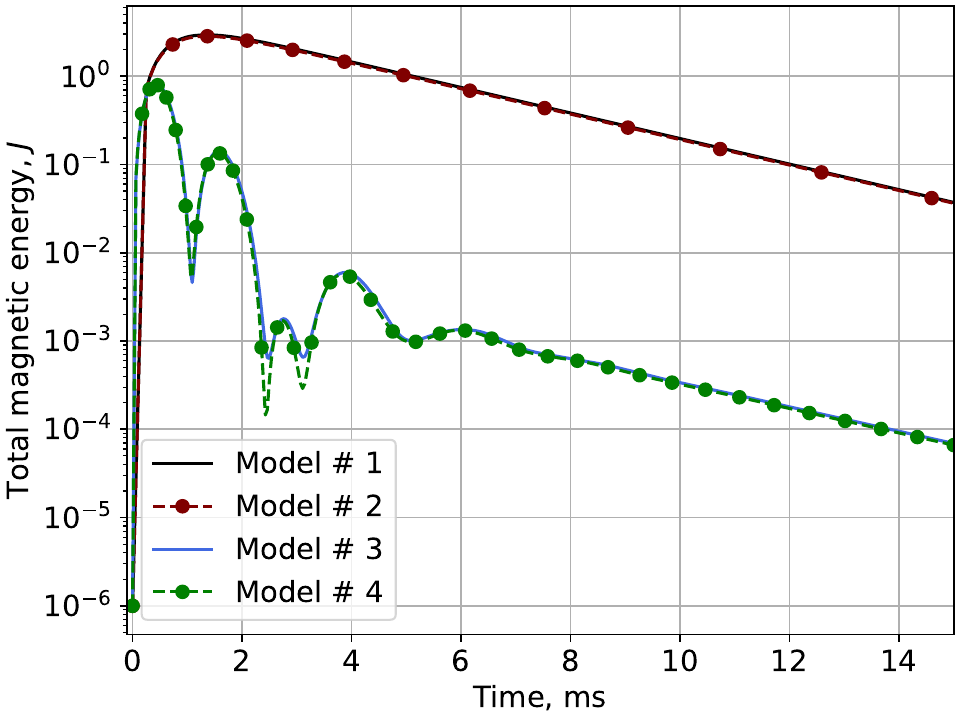}
    \caption{Energy of induced magnetic flux $W_m^i$  \eqref{eq:me} computed at $\alpha=0$ using four models listed in table \ref{table_model}.}
    \label{fig:Wmi-comp}
\end{figure}

\begin{figure}
    \centering
    \includegraphics[width=1.0\linewidth]{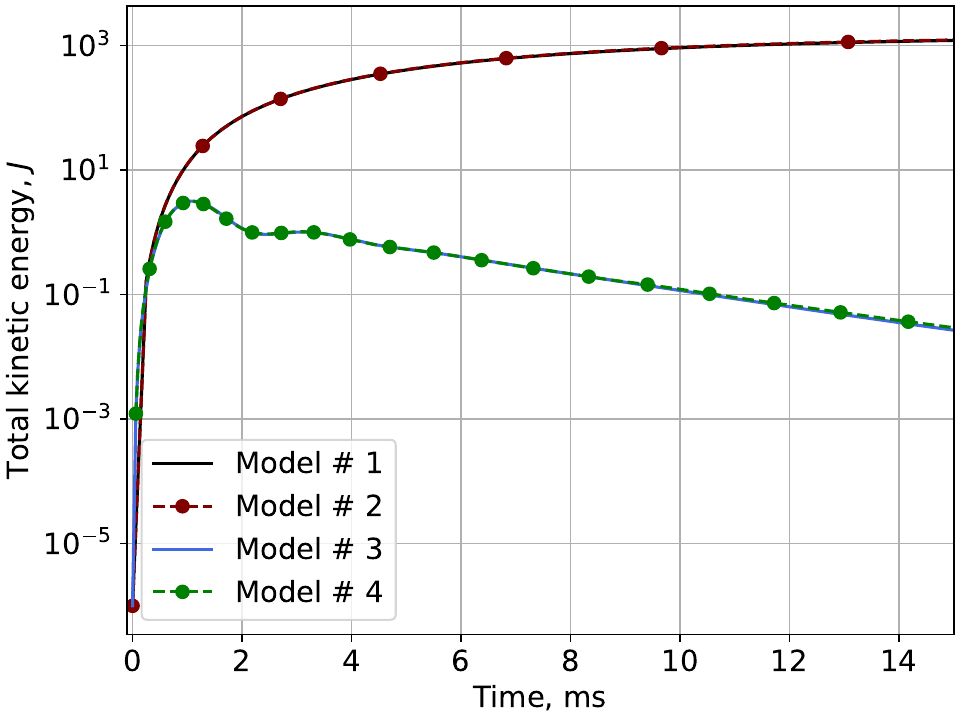}
    \caption{Kinetic energy  $W_k$  \eqref{eq:ke}} computed at $\alpha=0$ using four models listed in table \ref{table_model}.
    \label{fig:Wk-comp}
\end{figure}

\begin{figure}
    \centering
    \includegraphics[width=1.0\linewidth]{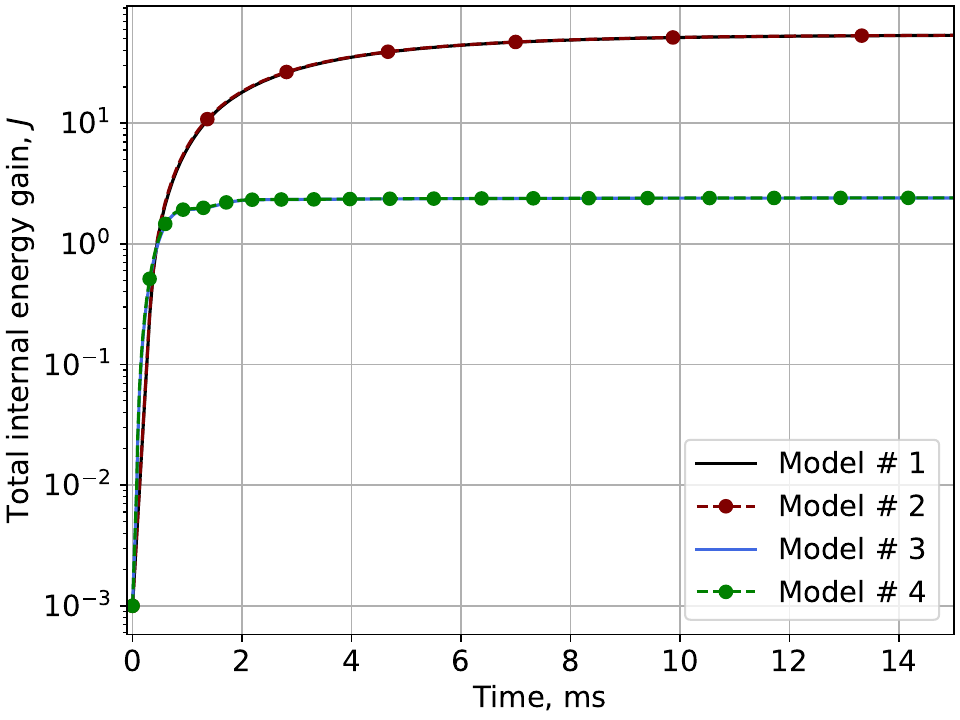}
    \caption{Gain of total internal energy  $W_i(t)$  \eqref{eq:ie}} computed at $\alpha=0$ using four models listed in table \ref{table_model}.
    \label{fig:Wi-comp}
\end{figure}

\begin{figure}
    \centering
    \includegraphics[width=1.0\linewidth]{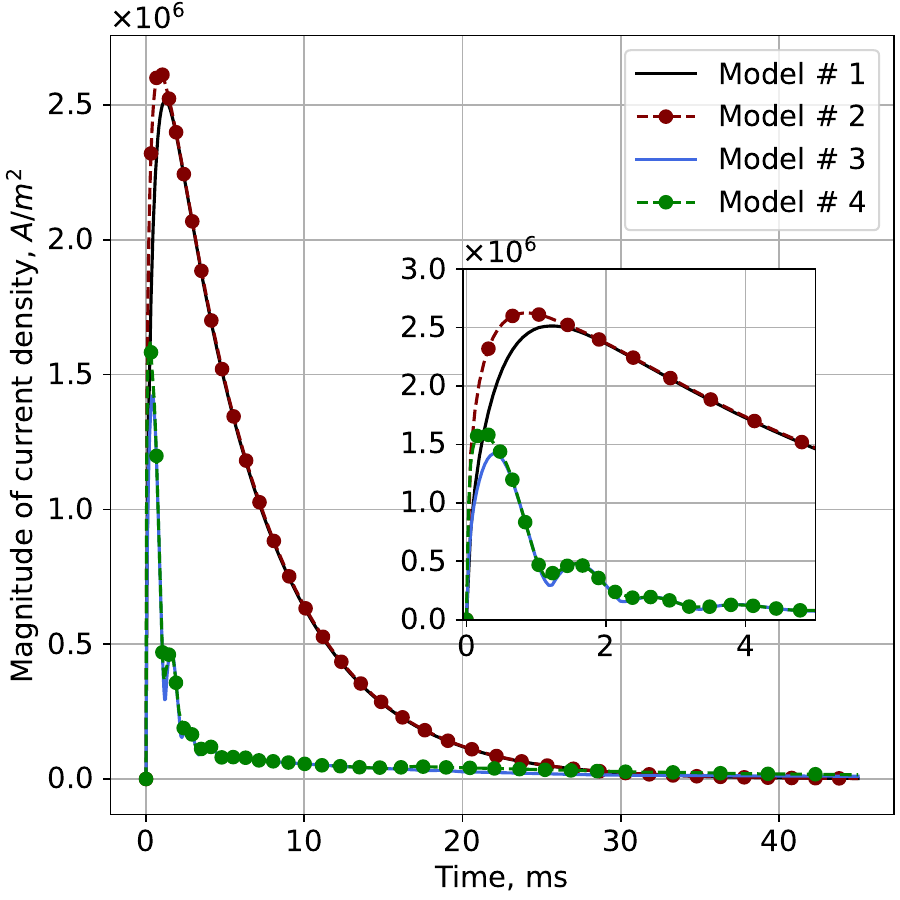}
    \caption{Maximum amplitude of electric current $|\mathbf{J}|_{max}$ in the solutions at $\alpha=0$ obtained using four models listed in table \ref{table_model}.}
    \label{fig:maxJ-comp}
\end{figure}

\begin{figure}
    \centering
    \includegraphics[width=1.0\linewidth]{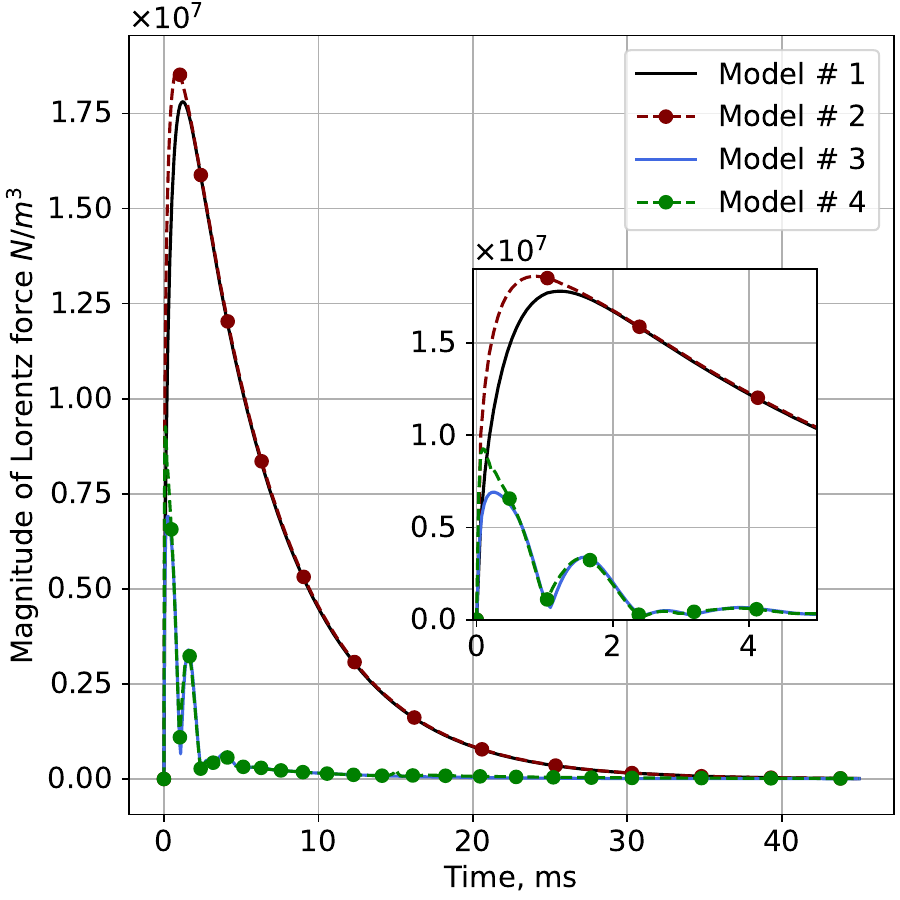}
    \caption{Maximum amplitude of Lorentz force $|\mathbf{F}|_{max}$ in the solutions at $\alpha=0$ obtained using four models listed in table \ref{table_model}.}
    \label{fig:maxF-comp}
\end{figure}

Comparison between the predictions of models 1-4 (see table \ref{table_model}) is presented. The results obtained for the poloidally oriented cavity with $\alpha=0$ (see figure \ref{fig:Sketch}) are shown. Figures \ref{fig:Wmi-comp}, \ref{fig:Wk-comp}, and \ref{fig:Wi-comp} illustrate the evolution of the energies $W_m^i$, $W_k$, $W_i$ generated in the cavity during the event. The maxima over the interior domain of the amplitudes of induced electric currents $|\mathbf{J}|_{max}$ and Lorentz force $|\mathbf{F}|_{max}$, the knowledge of which is essential for the assessment of the commonly made assumptions discussed in section \ref{sec:models}, are plotted in figures \ref{fig:maxJ-comp}
 and \ref{fig:maxF-comp}. 


We see that adding the exterior computational domain has only a modest effect on the solution. Switching from model 1 to model 2 or from model 3 to model 4 does not lead to a noticeable change of the kinetic and internal energies (see figures \ref{fig:Wk-comp} and \ref{fig:Wi-comp}). In the case of models 1 and 2, the same is true for the energy of the induced magnetic field (see figure \ref{fig:Wmi-comp}). Some difference between the curves of $W_m^i$ for models 3 and 4 is visible within a limited time interval, but this does not involve a significant change of the typical magnitude and behavior.  The curves of $|\mathbf{J}|_{max}$ and $|\mathbf{F}|_{max}$ in figures \ref{fig:maxJ-comp} and \ref{fig:maxF-comp} similarly demonstrate only a small quantitative change and no qualitative change of behavior when model 1 is replaced by model 2 or model 3 is replaced by model 4.

An explanation why the extension of the computational domain has only a modest effect on the solution will become evident section \ref{sec:res_a0}, where we illustrate the distributions of $\mathbf{b}$. We will see that the pseudovacuum boundary condition \eqref{eq:PSVBC} is certainly incorrect if applied at the boundary of the interior domain. The field lines of $\mathbf{b}$ penetrate the boundary at nonnormal angles and with nonzero gradients of the normal component. At the same time, the amplitude of $\mathbf{b}$ is small outside the fluid cavity and decreases rapidly with distance from it. As clearly demonstrated by the data in figures \ref{fig:Wmi-comp}-\ref{fig:maxF-comp}, the effect of $\mathbf{b}$ induced outside the fluid domain on the solution within it is weak.

The situation with the effect of the fluid velocity $\mathbf{u}$ in Ohm's law \eqref{eq:ohm} is entirely different. We see in figures \ref{fig:Wmi-comp}-\ref{fig:maxF-comp} that taking $\mathbf{u}$ into account, i.e., transitioning from model 1 to model 2 or from model 3 to model 4 profoundly changes the evolution of the system. Only during the initial period, approximately at $t<0.4$ ms, all predicted characteristics of the response remain nearly the same. At the end of this interval, the kinetic energy $W_k$ grows to about 2 J (see figure \ref{fig:Wk-comp}), which corresponds to the average rms velocity $U=(2V_I^{-1}\rho^{-1}W_k)^{1/2}\approx 0.2$ m/s (as we will see below, the spatial distributions of $\mathbf{F}$ and $\mathbf{u}$ are quite nonuniform, so significantly higher values of $|\mathbf{u}|$ are achieved locally).  Beyond this interval, the non-negligible $\mathbf{u}$ in Ohm's law strongly affects the dynamics. It causes a rapid drop and time oscillations of the induced electric current, magnetic field, and Lorentz force (see figures \ref{fig:Wmi-comp}, \ref{fig:maxJ-comp}, \ref{fig:maxF-comp}). Interestingly, the kinetic energy $W_k$ also decreases (see figure \ref{fig:Wk-comp}), which can only be caused in our system by the Lorentz force acting against the velocity (see \eqref{eq:balke}, \eqref{eq:ql}). This behavior is discussed in detail below. At this moment, we note that the simplified models 1 and 2, which ignore the generation of electric currents by fluid motion, grossly overestimate the velocity. The kinetic energy $W_k$ continues to grow and, by $t=15$ ms, exceeds 1200 J, which corresponds to, as we now understand, unrealistic $U\approx 5$ m/s.

The rapid order-of-magnitude decrease of the current magnitude and, respectively, the Joule dissipation means that the internal energy $W_i$ remains nearly constant after the initial growth (see figure \ref{fig:Wi-comp}).

\subsection{Transient response at $\alpha=0$ }\label{sec:res_a0}
A detailed description of the transient response, as it appears in our simulations using model 4, is given in this section. The results obtained for the poloidally oriented cuboid with $\alpha=0$ (see figure \ref{fig:Sketch}) are presented. The data obtained for a cuboid with its axis inclined at $\alpha=15^{\circ}$ with respect to the poloidal field $\mathbf{B}_p(t)$ are also shown in the figures, but discussed in section \ref{sec:res_a15}.

\begin{figure}
    \centering
    \includegraphics[width=1.0\linewidth]{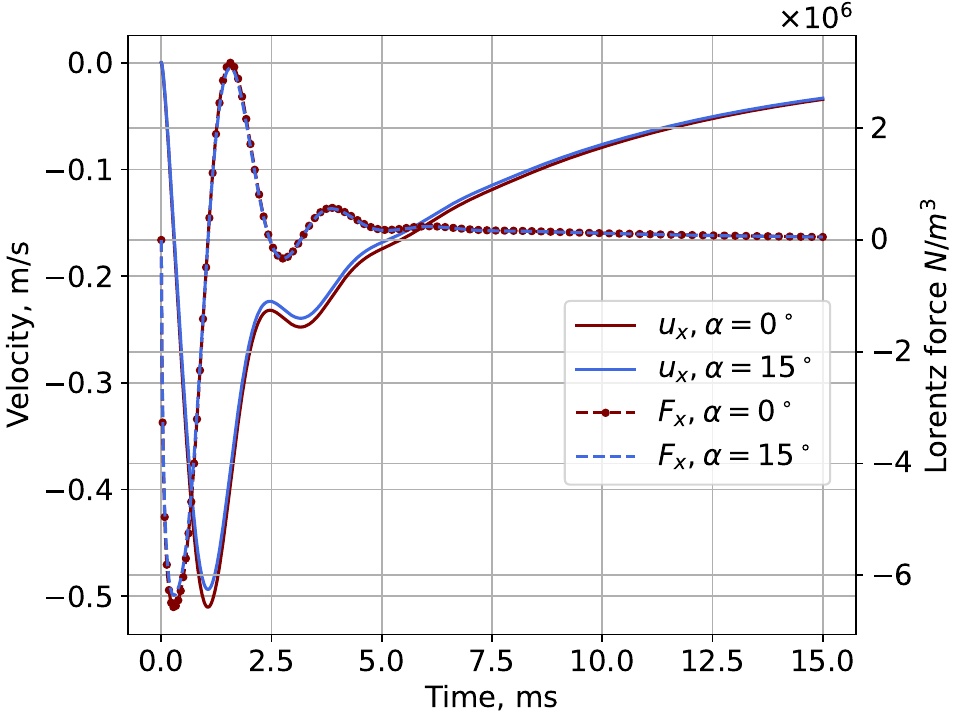}
    \caption{Values of $u_x$ and $F_x$ at the point (0, 0.49, 0). Results obtained for $\alpha=0$ and $\alpha=15^{\circ}$ are shown. }
    \label{fig:A}
\end{figure}

\begin{figure}
    \centering
    \includegraphics[width=1.0\linewidth]{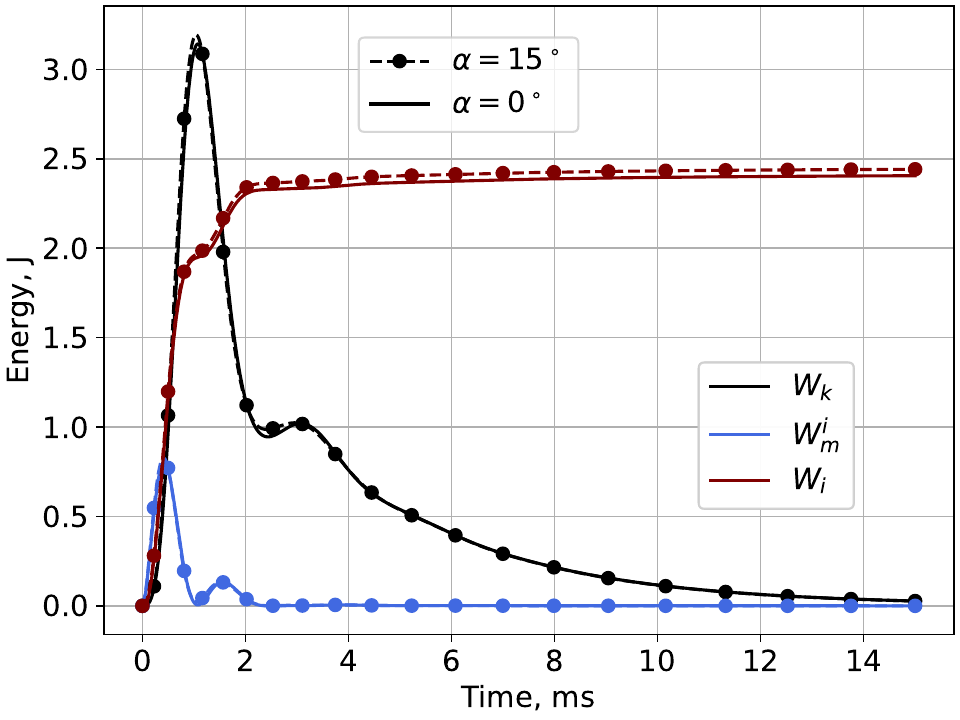}
    \caption{Total energies $W_k$, $W_i$, $W_m^i$ (see \eqref{eq:ke}, \eqref{eq:ie}, \eqref{eq:me})  shown as functions of time $t$ for $\alpha=0$ and $\alpha=15^{\circ}$. }
    \label{fig:B}
\end{figure}

\begin{figure}
    \centering
    \includegraphics[width=1.0\linewidth]{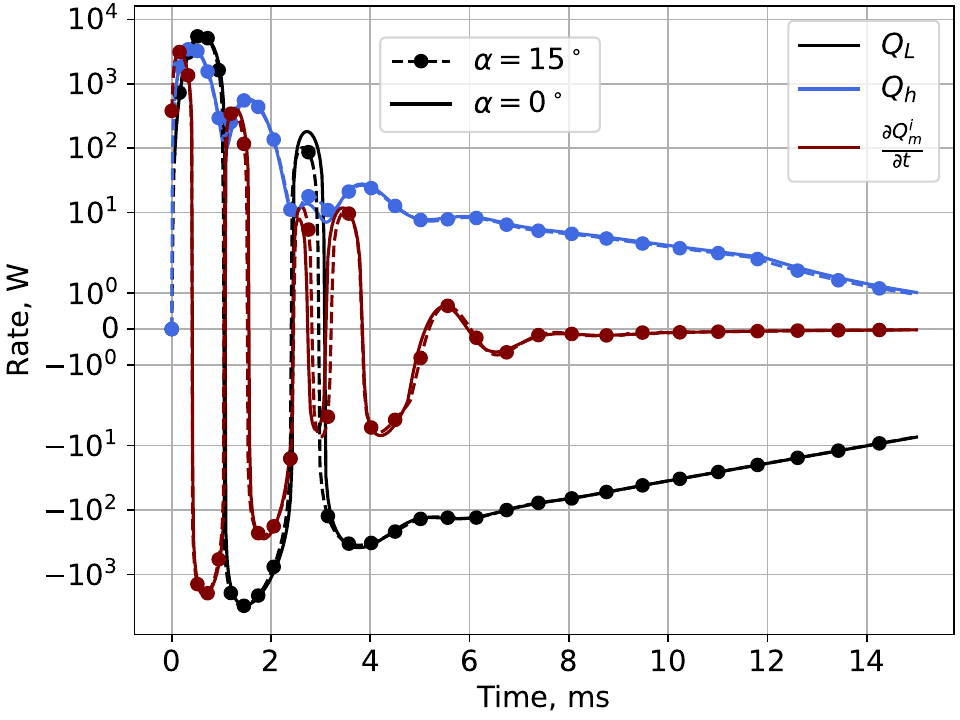}
    \caption{Energy transfer rates \eqref{eq:balme}-\eqref{eq:balie} shown as functions of time t: rate of mechanical work by the Lorentz force $Q_L$ (see \eqref{eq:ql}), rate of Joule dissipation of induced electric currents $Q_h$ (see \eqref{eq:qh}), and the rate of change of the induced component of the magnetic energy $dW_m^i/dt$ (see \eqref{eq:me}). Results obtained for $\alpha=0$ and $\alpha=15^{\circ}$ are shown. }
    \label{fig:C}
\end{figure}

\begin{figure*}
    \centering
    \includegraphics[width=1.0\linewidth]{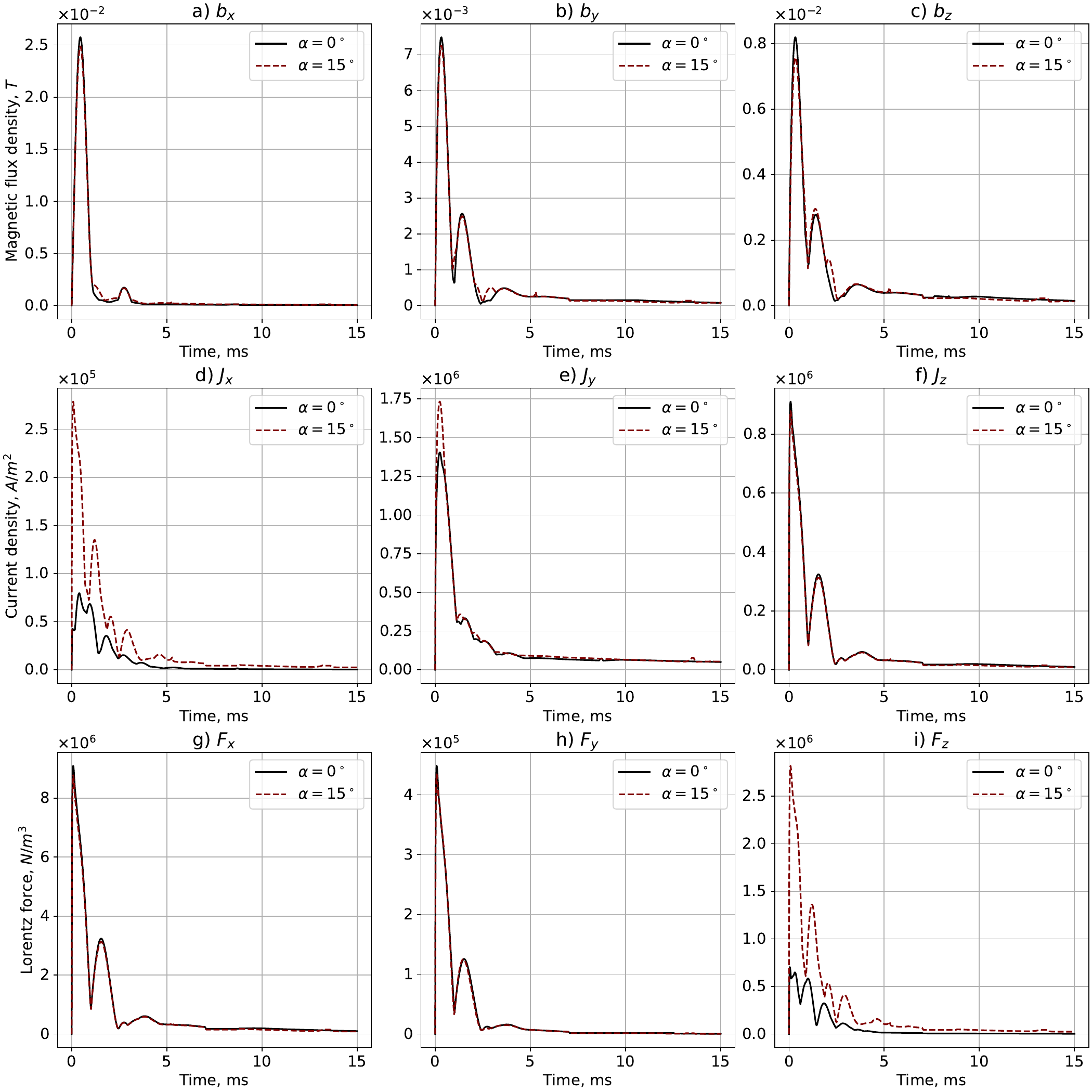}
    \caption{Temporal evolution of the maximum over the domain magnitude values of the components of the induced magnetic field $\mathbf{b}$ (top row), induced current desity $\mathbf{J}$ (middle row), and Lorentz force $\mathbf{F}$ (bottom row). Results obtained for $\alpha=0$ and $\alpha=15^{\circ}$ are shown.}
    \label{fig:D}
\end{figure*}

\begin{figure*}
    \centering
    \includegraphics[width=1.0\linewidth]{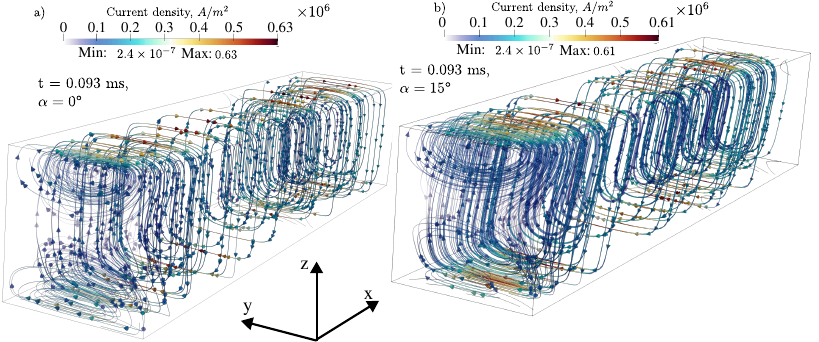}
    \caption{Distributions of the induced current density $\mathbf{J}$ at $t=0.093$ ms for $\alpha=0$ and $\alpha=15^{\circ}$. Distributions of the current magnitude $|\mathbf{J}|$ in cross-sections $x=const$ and current lines colored according to $|\mathbf{J}|$  are shown. Only the half of the domain with $0\le x\le 0.5$ m is shown for better visibility. }
    \label{fig:E}
\end{figure*}

\begin{figure*}
    \centering
    \includegraphics[width=1.0\linewidth]{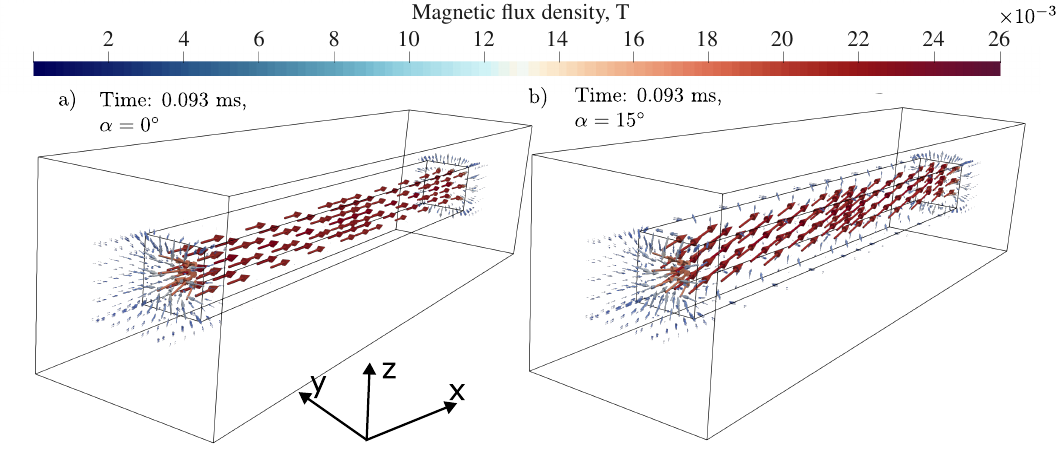}
    \caption{Distributions of the induced magnetic field $\mathbf{b}$ at $t=0.093$ ms. Vectors are colored  {and scaled} by the magnitude of $|\mathbf{b}|$.  {Magnetic flux density displayed on a uniformly spaced virtual grid comprising 1500 sample points.} Results obtained for $\alpha=0$ and $\alpha=15^{\circ}$ are shown. 
    }
    \label{fig:F}
\end{figure*}

\begin{figure}
    \centering
    \includegraphics[width=1.0\linewidth]{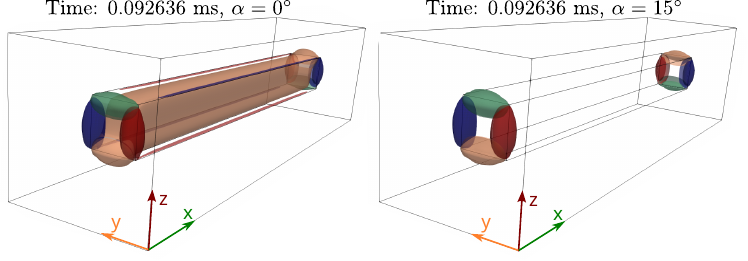}
    \caption{Iso-surfaces of the $y$ and $z$ components of the induced magnetic $\mathbf{b}$ at $t=0.093$ ms: $b_y=\pm 1$ mT (red and blue) and $b_z=\pm 1$ mT (orange and green). Results obtained for $\alpha=0$ and $\alpha=15^{\circ}$ are shown.}
    \label{fig:G}
\end{figure}

\begin{figure*}
    \centering
    \includegraphics[width=1.0\linewidth]{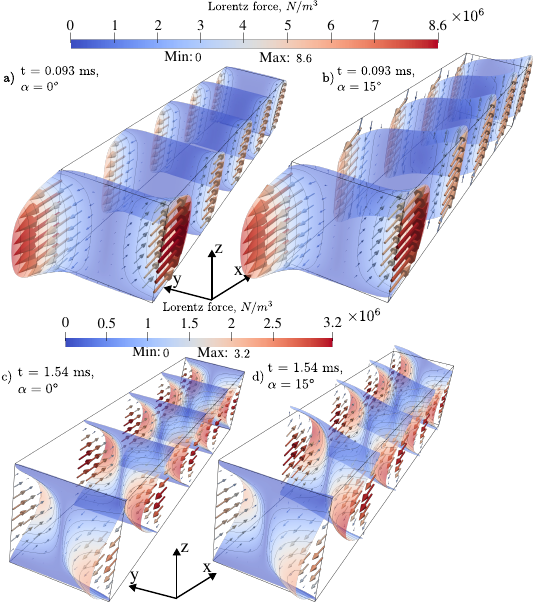}
    \caption{Distributions of the Lorentz force $\mathbf{F}$ at $t=0.093$ ms (top row) and $t=1.54$ ms (bottom row).  Vectors are colored by the magnitude of $|\mathbf{F}|$.  {The distributions are shown in five cross-sections $x=const$. Semi-transparent surfaces wrapping the vector and colored by the magnitude of $|\mathbf{F}|$ are  shown for better visibility.} Results obtained for $\alpha=0$ and $\alpha=15^{\circ}$ are shown. Only the half of the domain with $0 \le x \le 0.5 $ m is depicted.}
    \label{fig:H}
\end{figure*}

\begin{figure*}
    \centering
    \includegraphics[width=1.0\linewidth]{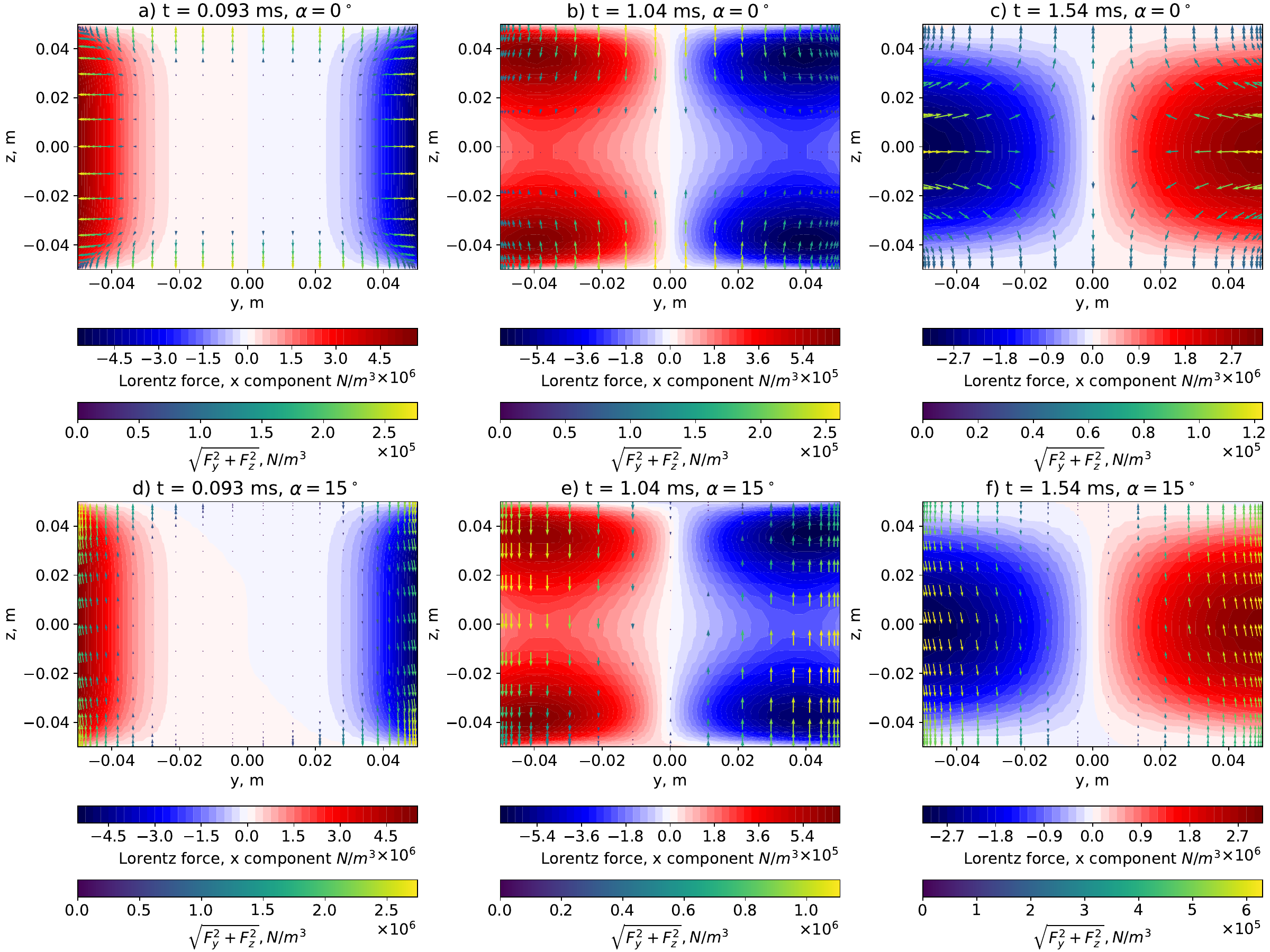}
    \caption{Distirbutions of the Lorentz force $\mathbf{F}$ in the middle cross-section $x=0.5$ m at $t\approx 0.093$ ms (left-hand side column), $t\approx 1.04$ (middle column) and $t\approx 1.54$ ms (right-hand side column). Continuous distributions of $F_x$ (red and blue colors) and vectors $(F_y,F_z)$ colored by their magnitude are shown. Results obtained for $\alpha=0$ (top row) and $\alpha=15^{\circ}$ (bottom row) are shown.} 
    \label{fig:I}
\end{figure*}

\begin{figure*}
    \centering
    \includegraphics[width=1.0\linewidth]{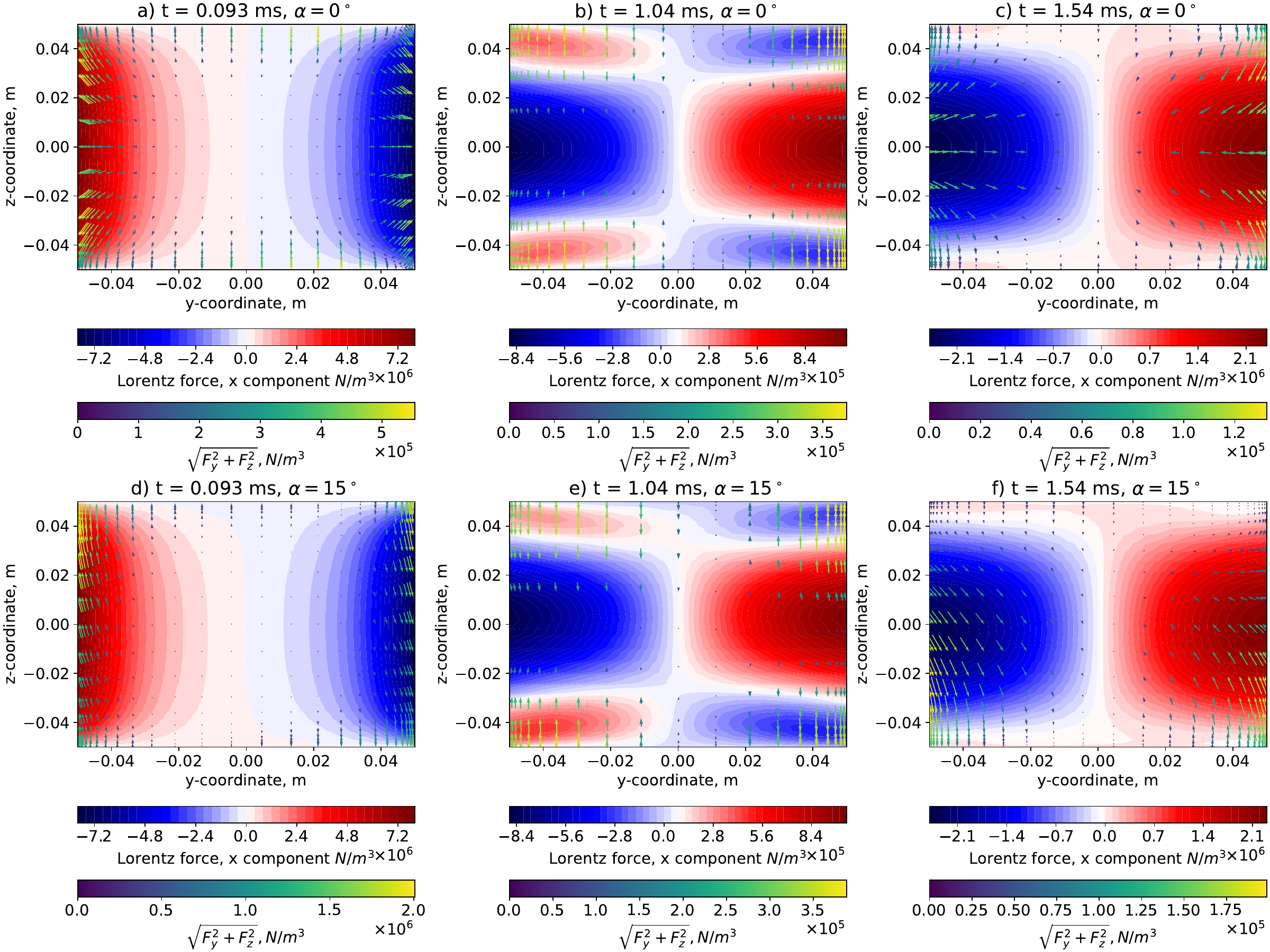}
    \caption{Distirbutions of the Lorentz force $\mathbf{F}$ in the cross-section $x=0.01$ m shown as in figure \ref{fig:I}.} 
    \label{fig:J}
\end{figure*}

\begin{figure*}
    \centering
    \includegraphics[width=1.0\linewidth]{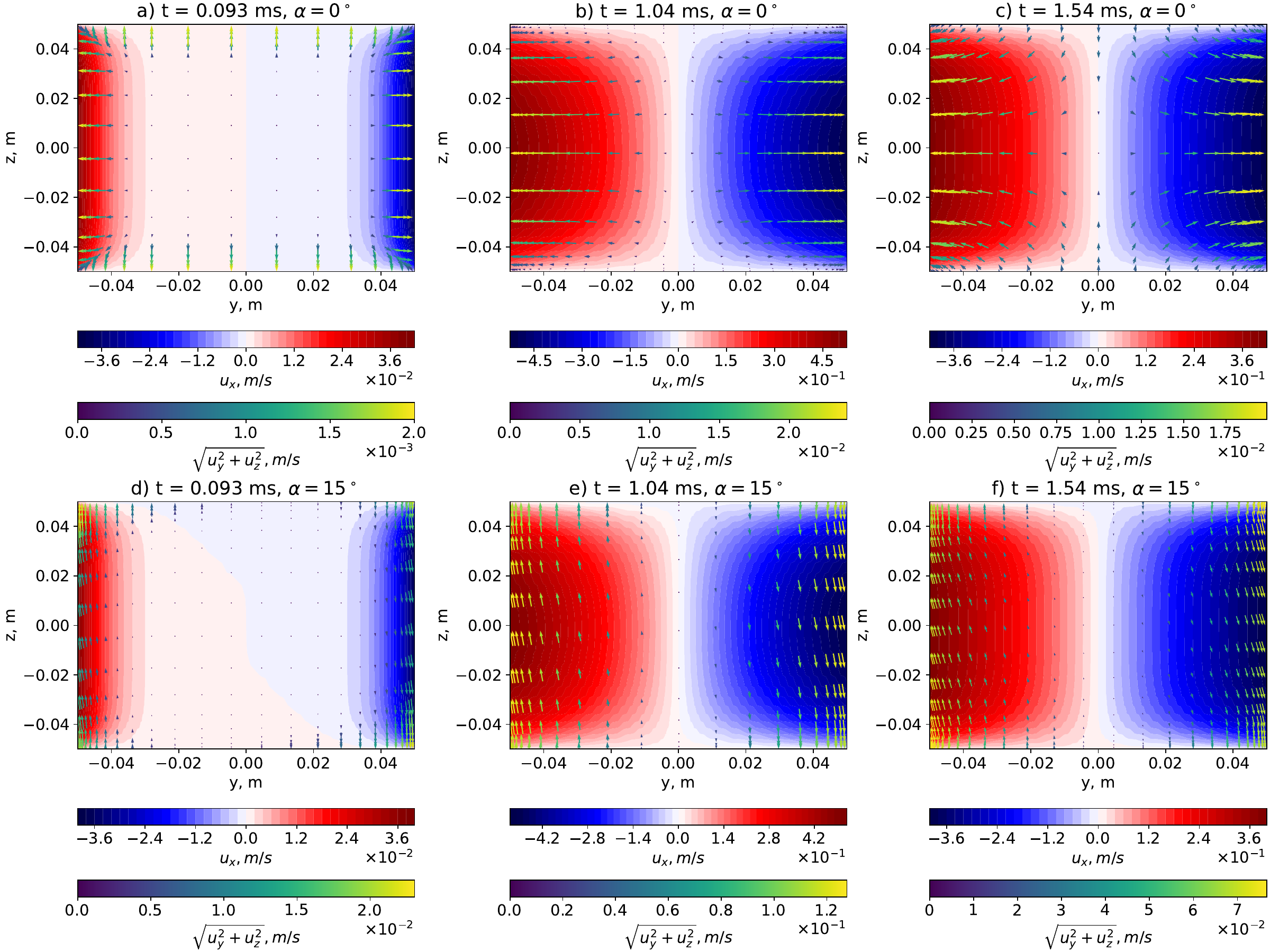}
    \caption{Distirbutions of the velocity $\mathbf{u}$ in the middle cross-section $x=0.5$ m at $t\approx 0.093$ ms (left-hand side column), $t\approx 1.04$ (middle column) and $t\approx 1.54$ ms (right-hand side column). Continuous distributions of $u_x$ (red and blue colors) and vectors $(u_y,u_z)$ colored by their magnitude are shown. Results obtained for $\alpha=0$ (top row) and $\alpha=15^{\circ}$ (bottom row) are shown.} 
    \label{fig:K}
\end{figure*}

The solution is illustrated using the time evolution of local and global characteristics (see figures \ref{fig:A}-\ref{fig:D}) and snapshots of three-dimensional and two-dimensional distributions of solution fields at various time points (see figures \ref{fig:E}-\ref{fig:K}).

As we will explain in detail shortly, two distinct stages can be identified: the initial stage, during which the fluid velocity does not have a significant impact, and the second stage, during which the evolution is strongly affected by the velocity. The separation between the two stages is inherently impresize, but we can estimate the threshold time $t_{crit}$ using two criteria. One is that models 1 and 2 should give nearly the same results as models 3 and 4 during the initial stage but show significant differences afterwards. As we discussed in section \ref{sec:mod_comp}, the data in figures \ref{fig:Wmi-comp}-\ref{fig:maxF-comp} suggest $t_{crit}\approx 0.4$ ms. 

Another criterion is based on the analysis of the flow dynamics provided in the following. We will see that during the initial stage, the amplitudes of $\mathbf{J}$, $\mathbf{F}$, $\mathbf{b}$ and $\mathbf{u}$ grow continuously. During the second stage, the Lorentz force may be directed in a direction opposite to the fluid velocity, causing a reduction in the induced fields. Consequently, $t_{crit}$ can be associated with the time of the first peak on the curves that represents the amplitudes of various perturbation fields. Here, different characteristics give different times. For the electric current, we see $t_{crit} \approx 0.093$ ms for $J_z$, larger $t_{crit} \approx 0.24$ ms for $J_y$ and $|\mathbf{J}|$, and $t_{crit} \approx 0.4$ ms for $J_x$ (see figures \ref{fig:D} and \ref{fig:maxJ-comp}). All components and the magnitude of the Lorentz force show $t_{crit}\approx 0.093$ ms (see figures \ref{fig:A}, \ref{fig:D} and \ref{fig:maxF-comp}). Values of $t_{crit}$ between 0.31 and 0.46 ms are found for the components and magnitude of $\mathbf{b}$ and the magnetic energy $W_m^i$ (see figures \ref{fig:B} and \ref{fig:D}). The peak of velocity $\mathbf{u}$ and the velocity-derived $W_k$ and $Q_L$ is delayed to $t_{crit}\approx 1.1$ ms (see figures \ref{fig:A}, \ref{fig:B} and \ref{fig:C}). Such different estimates are evidently due to the effects of the diffusion of the magnetic field and inertia of the fluid. A more detailed explanation will emerge in the following discussion.

The behavior observed during the initial stage is illustrated by the snapshots of the solution fields $\mathbf{J}$, $\mathbf{b}$, $\mathbf{F}$, and $\mathbf{u}$ at $t=0.093$ ms (see figures \ref{fig:E}a, \ref{fig:F}a, \ref{fig:G}a, \ref{fig:H}a, \ref{fig:I}a, \ref{fig:J}a, and \ref{fig:K}a). The behavior is open to a relatively simple interpretation. In the absence of significant fluid velocity, the varying imposed magnetic field $-(dB_p/dt)\mathbf{e}_x$ creates electric currents $(J_y,J_z)$ with clockwise (with respect to positive $x$) loops (see \eqref{eq:ampere}, \eqref{eq:ohm} and figures \ref{fig:D}d-f and \ref{fig:E}a). The induced magnetic field is predominantly in the component $b_x(y,z)$, although $\mathbf{b}$ becomes significantly three-component and three-dimensional near the ends $x=0$, $L_1$ of the fluid domain (see figures \ref{fig:D}a-c, \ref{fig:F}a and \ref{fig:G}a). The growth of $\mathbf{b}$ affects the distribution of $\mathbf{J}$ near the ends, which becomes increasingly dependent on $x$ and acquires a weaker but non-negligible third component $J_x$.

The distribution of the Lorentz force $\mathbf{F}$ during the initial stage is illustrated in figures \ref{fig:H}a, \ref{fig:I}a and \ref{fig:J}a. The dominant contribution is $\mathbf{J}\times B_t\mathbf{e}_y\approx -J_zB_t\mathbf{e}_x$. It is visible as zones of strong positive and negative values of $F_x$ near, respectively, the walls at $y=-W_1/2$ and $y=W_1/2$. The weaker contribution $\mathbf{J}\times B_p(t)\mathbf{e}_x\approx B_pJ_z\mathbf{e}_y-B_pJ_y\mathbf{e}_z$ has the form of a two-dimensional force field pushing the fluid particles toward the sidewalls (see figure \ref{fig:I}a). In the above expressions, the approximate equality is used because the weaker current component $J_x$ is ignored. Its effect, as well as the effect of $x$-variation of $\mathbf{J}$ leads to three-dimensional distortions of the force field, especially near the ends $x=0$, $L_1$ (see figures \ref{fig:H}a and \ref{fig:J}a).

The dominant force component $F_x$ accelerates the fluid in the positive $x$ direction near the wall $y=-W_1/2$ and in the negative $x$ direction near the wall at $y=W_1/2$ (see figure \ref{fig:K}a). The velocity $u_x$ increases significantly, for example, to about 0.5 m/s in the areas near the sidewalls at $t\approx 1$ ms (see figures \ref{fig:A} and \ref{fig:K}b). This triggers the end of the initial growth stage and the onset of the second stage of the evolution. The key factor is the contribution $\sigma u_x\mathbf{e}_x\times B_t\mathbf{e}_y=\sigma u_xB_t\mathbf{e}_z$ to Ohm's law. It is opposite to the contribution $\sigma E_{i,z}$ generated directly by $dB_p/dt$. Recalling that the $z$ component of $\mathbf{J}$ is the main contributor to the Lorentz force, we see how $\mathbf{F}$ is affected by $\mathbf{u}$. The new contribution $-\sigma u_x B_t^2 \mathbf{e}_x$ decelerates the fluid particles that are set in motion during the initial stage. At the same time, the accelerating force $-\sigma E_{i,z}B_t \mathbf{e}_x$ decreases as $E_{i,z}$ decreases following the decay of $dB_p/dt$. At some moment, the entire Lorentz force component $F_x$ starts to act against the motion of fluid particles. A proof that this actually happens is provided by the data in figure \ref{fig:C}. The work rate of the Lorentz force $Q_L$ becomes negative at $t\approx 1.1$ ms.

The solution fields during the second stage are illustrated by the snapshots of $\mathbf{F}$ and $\mathbf{u}$ at $t=1.54$ ms (see figures \ref{fig:H}c, \ref{fig:I}c, \ref{fig:J}c and \ref{fig:K}c).  The transformation is more complex than described above due to the three-dimensionality effects, Lorentz forces resulting from the interaction of currents with $B_p$, and the effect of the ends of the domain. Nontheless, the computed fields are fully consistent with the outlined scenario of the fluid velocity causing an opposing Lorentz force and a reversal of the initial trend of the system evolution. We see in figure \ref{fig:A} that the amplitudes of point values of $F_x$ and, with a delay, of $u_x$ start to decrease with time. The maxima of all components of $\mathbf{J}$, $\mathbf{b}$, and $\mathbf{F}$ also begin to decay (see figure \ref{fig:D}). A clear illustration of the transformation leading to this behavior is provided by the two-dimensional distributions of the components of $\mathbf{F}$ in figures \ref{fig:I}a-c (the middle cross-section $x=0.5$ m) and figures \ref{fig:J}a-c (the cross-section $x=0.01$ m near the end). We see the initial growth stage at $t=0.093$ ms in figures \ref{fig:I}a and \ref{fig:J}a. The transitional stage, during which $F_x$, while not yet reversed, is much weaker, is illustrated by the distributions at $t=1.04$ ms in figures \ref{fig:I}b and \ref{fig:J}b. The fully developed secondary stage, during which $F_x$ is reversed, is shown for $t=1.54$ ms in figures \ref{fig:I}c and \ref{fig:J}c. 

The final and most convincing proofs of the just introduced scenatrio are the decrease in the rate of Joule dissipation $Q_h$ and the rapid transition to negative values of the work rate by the Lorentz force $Q_L$ shown in figure \ref{fig:C}.

The consequent evolution of the system can be explained as a repeated, with modifications, realization of the scenario described above. After the fluid velocity is reduced in the course of the first force reversal (see the drop in the local value of $|u_x|$ and the kinetic energy $W_k$ at $1\lesssim t\lesssim 2$ ms in figures \ref{fig:A} and \ref{fig:B}), the cumulative Lorentz force $\mathbf{F}$ starts to accelerate the fluid particles in a way similar to that during the initial stage. $Q_L$ becomes positive (see figure \ref{fig:C}), $|u_x|$ and $W_k$ increase slightly (see figures \ref{fig:A} and \ref{fig:B}), and the maxima of the components of $\mathbf{b}$, $\mathbf{J}$, and $\mathbf{F}$ increase (see figures \ref{fig:maxJ-comp}, \ref{fig:maxF-comp}, \ref{fig:D}). This leads to the second force reversal, which occurs at a significantly lower average amplitude of $\mathbf{u}$ than the first reversal, since the base accelerating force $-\sigma E_{i,z} B_t\mathbf{e}_x$ is weaker at this stage.

The tendency toward further repeats of the cycle is indicated by the data (see, e.g., figure \ref{fig:C}), but complete reversals are not realized because of the decrease of the magnitude of $dB_p/dt$, which is the driver of the process. Multiple reversals cannot be excluded for other parameters of the system, for example, at stronger $B_p$ and $B_t$.  

 {The scenario of the liquid metal response reported in this section can be compared with the $x$-independent solutions for infinite channel and duct at $dB_p/dt=const<0$ obtained in \cite{kawczynski:2018thesis,smolentsev:2025}. The solutions include the full hydrodynamics of a viscous incompressible fluid, but limit the possible flow states to single-component fields $F_x$, $u_x$, and $b_x$, as well as the electric currents in the $y-z$ plane. Asymptotic steady-state and unsteady solutions evolving to steady states are obtained.}

 {Despite the profoundly different problem statement, our solution shows a certain agreement with the results of \cite{kawczynski:2018thesis, smolentsev:2025}. Considering only the first stage of the response and ignoring the effects of the ends of the domain, we find in figures \ref{fig:F}, \ref{fig:H}, \ref{fig:I}, and \ref{fig:K} the distributions of $\bm{b}$, $\bm{F}$, and $\bm{u}$, which are qualitatively similar to the distributions predicted in \cite{kawczynski:2018thesis, smolentsev:2025}. In particular, the dominant components of the fields are the $x$-components, and these components vary slowly along $x$. This provides a degree of validation for the idealized models of \cite{kawczynski:2018thesis,smolentsev:2025}. There are also similarities in spatial distributions of $b_x$, $F_x$, and $u_x$, including the concentration of $b_x$ in the core of the fluid domain (see figure \ref{fig:F}) or the fluid accelerated in the opposite direction near the opposite walls (see figure \ref{fig:K}).} 

 {At the same time, there are major differences between our solution and the solutions of  \cite{kawczynski:2018thesis,smolentsev:2025}. One is the loss of three-dimensionality and end effects in \cite{kawczynski:2018thesis,smolentsev:2025}. Another, probably more important difference is that the compex transient dynamics developing at a rapid and time-dependent variation of $B_p$ is reproduced by our model, but cannot be reproduced by the models of \cite{kawczynski:2018thesis,smolentsev:2025}. A complete analysis of the process obviously requires a model that, at the same time, includes the full hydrodynamics and considers the three-dimensional unstable dynamics of the system.}    

\subsection{Transient response at $\alpha=15^{\circ}$ }\label{sec:res_a15}
We now consider the effect of imperfect alignment between the longer side of the domain and the poloidal magnetic field. As illustrated in figure \ref{fig:Sketch}, this can be described by a nonzero angle $\alpha$ between $\mathbf{B}_p$ and the coordinate $x$ (see \eqref{eq:B_ext}). The results obtained at $\alpha=15^{\circ}$ are presented in figures \ref{fig:A}-\ref{fig:K} side-by-side with the analogous results for $\alpha=0$.

The main difference between the solutions at $\alpha=0$ and $\alpha=15^{\circ}$ is that the quasi-two-component fields with $|J_x|\ll |J_y|, |J_z|$, $|b_x|\gg |b_y|, |b_z|$ and $|F_x|\gg |F_y|, |F_z|$ are not observed at $\alpha=15^{\circ}$ even at the initial stage. From the onset, the time-dependent imposed magnetic field $-dB_p/dt(\cos\alpha \mathbf{e}_x+\sin\alpha \mathbf{e_z})$ creates a field of current density $\mathbf{J}$ with all three significant components. This is illustrated in figure \ref{fig:E} and given quantitative evidence in figure \ref{fig:D}, where we see that the maximum amplitude of $J_x$ increases approximately four times greater than at $\alpha=0$ and becomes much closer to the maximum amplitudes of $J_y$ and $J_z$. The induced magnetic field $\mathbf{b}$ shows stronger components $b_y$ and $b_z$ outside the fluid domain (see figures \ref{fig:F} and \ref{fig:G}). For example, we see in figure~\ref{fig:G}b that the component $b_z$ of significant (1 mT) magnitude appears in the exterior domain over the entire length of the cuboid. 

Finally, and most importantly, the field of Lorentz force $\mathbf{F}$ is modified (see figures \ref{fig:H}b,d, \ref{fig:I}d-f and \ref{fig:J}d-f). A significant contribution $F_z=J_xB_t$ appears at both stages of evolution, both in the core of the domain (see figure \ref{fig:I}d-f) and near the ends (see figure \ref{fig:J}d-f). 

We also see in figures \ref{fig:I}d-f and \ref{fig:J}d-f that the force component $F_z$ experiences a transformation similar to that described in section \ref{sec:res_a0} for $F_x$. The direction of $F_z$ is reversed between $t=0.093$ ms (initial stage of the evolution) and $t=1.54$ ms (secondary stage). The only plausible explanation is the one proposed earlier for $F_x$. Acceleration by $F_z$ during the initial stage creates zones of flow with high velocity $u_z$ near the walls at $y=\pm W_1/2$. The resulting contribution to Ohm's law $J_x\approx -\sigma u_z B_t$ leads to a growing contribution to the Lorentz force $F_z\approx -\sigma u_zB^2_t$, which opposes the fluid motion. 

Interestingly, the clearly visible impact of $\alpha$ on the spatial structure of the solution fields is not accompanied by a comparably significant change in the quantitative characteristics of the system behavior. Close values of the point values of $u_x$ and $F_x$ (see figure \ref{fig:A}) and the maxima of the components of $\mathbf{J}$, $\mathbf{b}$, and $\mathbf{F}$ (see figure \ref{fig:D}) are found at $\alpha=0$ and $\alpha=15^{\circ}$. Only substantial differences are detected for $J_x$ and $F_z$, i.e., for the components with small contributions in the magnitudes $|\mathbf{J}|$ and $|\mathbf{F}|$. The total energies $W_k$, $W_m^i$, $W_i$ and the energy transformation rates $Q_L$, $Q_H$, $dW_m^i/dt$ are not significantly affected (see figures \ref{fig:B} and \ref{fig:C}). 

The spatial distributions in figures \ref{fig:E}-\ref{fig:K} also suggest that although a nonzero $\alpha$ introduces a significant third component in the vector fields, it does not lead to significant three-dimensionality. The fields remain varying only slowly with $x$ apart from the areas adjacent to the ends $x=0,L_1$ of the domain. 

In summary, we see that an imperfect polioidal orientation of the fluid domain, while affecting the spatial distributions of $\mathbf{J}$, $\mathbf{b}$ and $\mathbf{F}$, does not significantly affect the characteristics most important for us: the typical and maximum amplitudes of $\mathbf{J}$, $\mathbf{b}$, $\mathbf{F}$ and $\mathbf{u}$. Not less importantly, the general scenario of the system evolution, which consists of the initial growth, the first reversal, and the following oscillations superimposed on the decay of the perturbations, remains largely the same as in the case of the perfectly poloidal orientation. 
 
\section{Concluding remarks}\label{sec:conclusion}
We have completed an exploratory analysis of the processes that arise in a liquid-metal blanket of a fusion reactor in response to a plasma disruption event. An idealized cuboid geometry, an approximated time profile of the imposed magnetic field, and a simplified hydrodynamic model were used. In other major aspects, the analysis reproduced the actual blanket conditions. This, in particular, included the realistic strength and time scale of the imposed magnetic field, dimensions of a blanket module, and material properties of the fluid (PbLi at 573 K).

The order-of-magnitude analysis of section \ref{sec:models} provided preliminary estimates and presented a framework for analyzing the applicability of commonly used modeling assumptions. It was concluded that many such assumptions are always valid, but others, most significantly those of incompressibility and constant physical properties of the metal, may become invalid at certain quantitative characteristics of the response. 

The numerical simulations pursued two goals. One was to evaluate the effect of two major simplifications of a numerical model on the results. It was found that limiting the computational domain to the fluid interior and imposing the pseudo-vacuum conditions on its boundaries, while it is incorrect in principle, does not significantly change the solution. It must be stressed that the tests were conducted under the assumption of electrically nonconducting surroundings, and for the symmetric situation, in which the time-varying poloidal magnetic field was perfectly aligned with the axis of the cuboid. It is possible that the simplified model is significantly less accurate if one of these conditions is not satisfied. 

The other major simplification of the model, namely omitting the fluid velocity in the expression of Ohm's law for induced electric currents \eqref{eq:ohm} proved to be much more detrimental to accuracy. We have found that such a simplified model correctly captures the physics only during a very short initial period of the response, about 1 ms at the parameters explored in this paper. At the end of this period, the fluid accelerates to a moderate velocity (about 0.5 m/s), which, however, is sufficient to completely change the distribution of the electric currents. The Lorentz force starts acting against the flow. The subsequent evolution of the system is characterized by reversals of the force field superimposed on rapid decay of the current density, force, and velocity fields. The simplified model cannot capture the reversals and, as a result, predicts grossly overestimated amplitudes and slower decay of the fields.

The picture of the fluid response emerging from our study is still incomplete because it is based on the reduced hydrodynamic model \eqref{eq:velocity} and several other simplifications, such as an idealized geometry of a blanket module, the absence of its electrically conducting surroundings, and using the approximation \eqref{eq:B_ext} of the time variation of the magnetic field. Removing these simplifications will require advanced and large-scale simulations based on a detailed blanket design and a specific transient event. Such simulations will have to include the behaviors of the plasma and the surrounding reactor structure as coupled processes.  Nevertheless, the overall scenario of the liquid metal experiencing an initial growth of induced fields and then the evolution determined by rapidly changes and gradually decreasing in strenth acceleration/decceleration cycles caused by reversals of the Lorentz force appears to be robust and likely to be realized in actual blankets during actual transient plasma events. This has significant implications for the design and operation of blankets. We mention those that seem especially important at the current level of understanding.

A reevaluation of the estimates of the currents, forces, and fluid velocities that arise within the blanket appears to be necessary. Such estimates are often based on the order-of-magnitude analysis, which ignores the spatial distribution of the solution fields and roughly corresponds, in the scope of the physics it captures, to the initial stage of the evolution revealed by our analysis. The simulation results presented in this paper suggest that such an analysis can correctly predict the magnitudes of $\mathbf{b}$, $\mathbf{J}$, $\mathbf{F}$, and $\mathbf{u}$ during the short period ($\sim 1$ ms) at the beginning of the event. Application of the order-of-magnitude analysis to later stages of the blanket response results in severely overestimated magnitudes. 

In fact, the revealed response scenario can be used to assess the maximum velocity $u_{max}$, to which the metal can be accelerated from rest during a transient plasma event. $u_{max}$ approximately corresponds to the velocity at which the first reversal of the Lorentz force occurs, i.e. the velocity at which the dominant component of $\mathbf{u}\times \mathbf{B}$ in \eqref{eq:ohm} grows large enough to balance the corresponding component of $\mathbf{E}_i$. Based on the analysis in section \ref{sec:res_a0}, this means $u_{max}B_t \sim E_{i,z}$. Estimating $E_{i,z}\sim (W_1/2)dB_p/dt$ and $dB_p/dt\sim \tau_0^{-1} B_p^0$, we find 
\begin{equation}\label{eq:umax}
    u_{max}\sim \frac{W_1}{2}\frac{B_p^0}{\tau_0 B_t},
\end{equation}
where $W_1$ can be understood with the general meaning as the typical length scale of the system in the direction transverse to the variable component of the magnetic field.  Using the parameters of our system, for which $W_1=0.1$ m, $\tau_0=6\times 10^{-3}$ s, and $B_p^0/B_t=0.05$, we find, in good agreement with the simulation results,  $u_{max}\approx 0.4$ m/s.

The potential implications for hydrodynamic modeling were discussed in section \ref{sec:models}. We now return to this question using the simulation data. The preliminary estimates of the maximum current density $J\sim 10^7$ A/m$^2$ and force $F\sim 10^7$ N/m$^3$ are confirmed, although such high values are only approached during the initial stage. The assumptions of the negligible role of the Joule dissipation in the heat balance, constant physical property coefficients, and negligibly small typical density variations are shown to be valid for our model and can be plausibly assumed to be valid for actual systems with similar parameters. 

At the same time, and this appears to be a major conclusion of our work, the assumption of incompressibility of the fluid is cast in doubt by the simulation data. The typical time scale of the variation of the Lorentz force is not the time scale of the transient plasma event $\tau$, but the much smaller ($\sim 10^{-3}$ s) typical time between the start of the event and the first reversal of the force. This time scale is comparable to the typical time $\tau_w\sim 5\times 10^{-4}$ s of the propagation of pressure waves across the fluid domain. The significant effect of pressure waves on the fluid behavior is anticipated. This requires the use of a compressible fluid model in the hydrodynamic analysis if a detailed and accurate picture of the transient response of the blanket is desired.

\acknowledgments{The work of Oleg Zikanov was done under the subcontract CW52278 between the University of Michigan - Dearborn and the Oak Ridge National Laboratory as a part of the SciDAC project ``Center for Simulation of Plasma - Liquid Metal Interactions in Plasma Facing Components and Breeding Blankets of a Fusion Power Reactor'' supported by the U.S. Department of Energy under the contract DE-AC05-00OR22725.}

\providecommand{\noopsort}[1]{}\providecommand{\singleletter}[1]{#1}%

\end{document}